\documentstyle[11pt,aaspp4]{article}

\lefthead{Hogerheijde et~al.}
\righthead{Tracing the envelopes around embedded low-mass YSOs}
\begin{document}

%----------------------------------------------------------------------

%%% 1. TITLE PAGE

\bigskip
\title{Tracing the envelopes around embedded low-mass young stellar
objects with HCO$^+$ and millimeter-continuum observations}

\author{Michiel R. Hogerheijde and Ewine F. van~Dishoeck}
\affil{Sterrewacht Leiden, P.O. Box~9513, 2300~RA Leiden, The
Netherlands}
\authoraddr{Sterrewacht Leiden, P.O. Box~9513, 2300~RA Leiden, The
Netherlands}
\author{Geoffrey A. Blake}
\affil{Division of Geological and Planetary Sciences, California
Institute of Technology, MS~150--21, Pasadena, CA~91125} 
\and
\author{Huib Jan van~Langevelde}
\affil{Joint Institute for VLBI in Europe, P.O. Box~2, 7990~AA,
Dwingeloo, The Netherlands}

%----------------------------------------------------------------------

%%% 2. ABSTRACT AND SUBJECT HEADINGS

\begin{abstract} 

The envelopes and disks around embedded low-mass young stellar objects
(YSOs) are investigated through millimeter continuum and HCO$^+$ line
emission. Nine sources, selected on the basis of their HCO$^+$ 3--2
emission from an IRAS flux- and color-limited sample of 24 objects,
are observed in $\lambda=3.4$ and 2.7~mm continuum emission with the
Owens Valley Millimeter Array, and in the HCO$^+$ and H$^{13}$CO$^+$
4--3, 3--2, and 1--0 transitions at the James Clerk Maxwell and
IRAM~30m telescopes. All nine sources are detected at 3.4 and 2.7~mm
in the interferometer beam, with total fluxes between 4 and
200~mJy. The visibilities can be fit with an unresolved ($<3''$) point
source, and, in about half of the sources, with an extended
envelope. The point sources, presumably thermal dust emission from
circumstellar disks, typically contribute 30--75\% of the continuum
flux observed at 1.1~mm in a $19''$ beam, assuming a spectral slope
of~2.5.  The fact that at least two-thirds of our sources show
point-source emission indicates that circumstellar disks are
established early in the embedded phase.  The remainder of the 1.1~mm
single dish flux is attributed to an extended envelope, with a mass of
$0.001$--$0.26$~$M_{\sun}$ within a $19''$ beam. In HCO$^+$, the
$J$=1--0 line is seen to trace the surrounding cloud, while the
emission from $J$=3--2 and 4--3 is concentrated toward the sources. All
sources look marginally resolved in these lines, indicative of a
power-law brightness distribution. A beam-averaged HCO$^+$ abundance
of~$(1.2\pm 0.4) \times 10^{-8}$ with respect to H$_2$ is derived. 

The 1.1~mm continuum fluxes and HCO$^+$ line intensities of the
envelopes correlate well, and are modeled with the simple inside-out
collapse model of Shu \markcite{shu77}(1977) and with power-law density
distributions of slopes $p=$~1--3. All models provide satisfactory fits
to the observations, indicating that HCO$^+$ is an excellent tracer of
the envelopes. Of the 15 sources of the original sample that were either
undetected in HCO$^+$ 3--2 or too weak to be selected, seven show 1.1~mm
single-dish fluxes comparable to our objects. It is proposed that all of
the 1.1~mm flux of the former sources should be attributed to compact
circumstellar disks. The relative evolutionary phase of a YSO, defined
as the current ratio of stellar mass over envelope mass, is traced by
the quantity $\int T_{\rm mb} dV ({\rm HCO^+}\,3\!-\!2) \,/\, L_{\rm
bol}$. Sources which are undetected in HCO$^+$ are found to to have
significantly lower values in this tracer than do the objects of our
subsample, indicating that the former objects are more evolved. The
sources which are weak in HCO$^+$ 3--2 are indistinguishable from our
subsample in this tracer, and have intrinsically low masses. It is
concluded that HCO$^+$, especially in its 3--2 and 4--3 transitions, is
a sensitive tracer of the early embedded phase of star formation.

\end{abstract}


% ----------------------------------------------------------------------

%%% 3. TEXT AND ACKNOWLEDGMENTS

\section{Introduction}

Low-mass protostars spend the earliest stages of their evolution
embedded in large amounts of gas and dust. These envelopes are
dispersed over the course of a few times~$10^5$ yr, after which the
pre--main-sequence object, possibly surrounded by an accretion disk or
its remnant, is revealed at near-infrared and optical
wavelengths. Even though the general characteristics of this evolution
are well understood (see Shu et~al.\ \markcite{shu93}1993 for a recent
overview), many details remain uncertain. During the early embedded
phase, various physical processes occur simultaneously within a few
thousand AU of the forming star.  These include the formation and
evolution of a circumstellar disk, accretion of matter onto the
protostar, the onset of an outflow, and the dissipation of the
envelope. The evolution of these phenomena is best studied by
observing a well-defined sample of low-mass young stellar objects
(YSOs), and by probing all components of their environment on the
relevant spatial scales.

Several surveys of the (sub-)~millimeter continuum emission of various
classes of YSOs have been undertaken in recent years using single-dish
telescopes (e.g., Beckwith et~al.\ \markcite{bec90}1990, Barsony \&
Kenyon \markcite{bar92}1992, Moriarty-Schieven et~al.\
\markcite{mor94}1994, Andr\'e \& Montmerle \markcite{and94}1994,
Osterloh \& Beckwith \markcite{ost95}1995). These studies have
established that $\sim 60\%$ of T~Tauri stars (class~II objects; Lada
\markcite{lad87}1987) have circumstellar disks, and that the mass of
these disks decreases over a period of a few times $10^6$~yr. The
interpretation of such data in the earlier embedded phase (class~I) is
much more ambiguous, since, in addition to the disks, the envelopes
also contribute to the observed single-dish flux. During the embedded
phase, matter is transferred from the envelope to the (growing)
disk. Since the latter is likely to be optically thick even at
millimeter wavelengths the spectral slope of its emission is
determined by geometry and orientation, and only lower limits of its
mass can be inferred. The high spatial resolution offered by
interferometer observations is essential to separate the relative
contributions of disk and envelope, as demonstrated by Keene \& Masson
\markcite{kee90}(1990), Terebey, Chandler \& Andr\'e
\markcite{ter93}(1993) and Butner, Natta \& Evans
\markcite{but94}(1994). Circumstellar disks have been tentatively
resolved for only a few embedded objects, with typical semi-major axes
of 60--80~AU (L1551~IRS~5, HL~Tau: Lay et~al.\ \markcite{lay94}1994;
T~Tau: van~Langevelde, van~Dishoeck \& Blake \markcite{hvl95}1995;
L1551~IRS~5: Mundy et~al.\ \markcite{lgm96}1996). Caution in the
interpretation of these limited interferometry data is urged, since
the recent results of Looney et~al.\ \markcite{loo96}(1996) show that
the allegedly resolved disk around L1551~IRS~5 is in fact a binary,
with each component surrounded by an unresolved disk.

An alternative approach for studying circumstellar envelopes is to use
spectral line observations to test models of cloud collapse such as
those developed by Shu \markcite{shu77}(1977), Terebey, Shu \& Cassen
\markcite{ter84}(1984), Galli \& Shu \markcite{gal93a}(1993), Fiedler
\& Mouschovias \markcite{fie92}(1992), \markcite{fie93}(1993), Boss
\markcite{bos93}(1993), and Foster \& Chevalier
\markcite{fos93}(1993).  Whereas the continuum data are only sensitive
to the total mass in the beam, line observations also test the density
and velocity structure. Detailed calculations of the molecular
excitation and radiative line transfer in model envelopes have been
performed by Zhou (\markcite{zho92}1992, \markcite{zho95}1995), Zhou
et~al.\ \markcite{zho93}(1993), Walker, Narayanan \& Boss
\markcite{wal94}(1994), and Choi et~al.\ \markcite{cho95}(1995), but
comparison with observations has been limited to a few specific, very
young class~0 objects such as B335 and IRAS~16293$-$2422. Although the
results seem to be consistent with the simple inside-out collapse
models of Shu \markcite{shu77}(1977) and Terebey et~al.\
\markcite{ter84}(1984), the models have not yet been tested over the
entire time span of the embedded phase. Ohashi et al.\
(\markcite{oha91}1991, \markcite{oha96a}1996) and Moriarty-Schieven
et~al.\ (\markcite{mor92}1992, \markcite{mor95}1995) have surveyed a
large sample of embedded objects using various transitions of the CS
molecule to probe a range of excitation conditions.  This molecular
approach works best if the abundance of the adopted species does not
change with time through, e.g., depletion, and if it uniquely traces
the envelope but none of the other components.  As will be shown in
this work, the HCO$^+$ ion may be better suited for this purpose than
CS.

There are several observational developments which make the detailed
study of a larger sample timely. First, millimeter interferometers
have expanded in size over the last few years, increasing both the
sensitivity and mapping speed by large factors.  Second, single-dish
submillimeter telescopes have been equipped with low-noise SIS
detectors at high frequencies, which makes observations of the less
massive, presumably more evolved, envelopes feasible. Third, powerful
bolometer arrays to trace the continuum emission from dust are just
coming on line.

Stimulated by the availability of these new techniques, we have
carried out a detailed study of nine embedded low-mass YSOs in the
nearby (140~pc) Taurus--Auriga star-forming region, using both single
dish and interferometric techniques. These objects span a large range
in luminosity and outflow activity, presumably reflecting variations
in age and mass. Both continuum emission and a variety of molecular
transitions have been observed, tracing densities from $10^4$ to
$10^8$~cm$^{-3}$, temperatures between 10 and 150~K, and angular
scales of $3''$ to 2$'$ ($400$~AU to 15,000~AU at 140~pc), independent
of orientation.  This large range in probed physical conditions and
scales, achieved through the combination of single-dish and
interferometer observations, distinguishes our work from most studies
mentioned above.

The aim of this paper, the first in a series, is to investigate the
evolution of the mass and density structure of the envelope using both
continuum and line observations. The results will be tested against
one of the simplest models of protostellar collapse (Shu
\markcite{shu77}1977), and closely related power-law density
distributions. A second objective is to investigate whether even the
most embedded YSOs are already surrounded by circumstellar disks.
Ohashi et~al.\ \markcite{oha96a}(1996) found that only a small
fraction, $\sim 15\%$, of the embedded objects shows compact 98~GHz
emission in the Nobeyama interferometer. Compared to the number of
T~Tauri stars with disks, this detection rate is low, and has been
interpreted as an indication of disk growth during the embedded
phase. A final objective is to study the usefulness of certain
molecules, especially HCO$^+$, as reliable tracers of the
circumstellar environment. This is important not only for constraining
the physical structure, but also for establishing a baseline for
future studies of the chemical evolution.

The choice of the Taurus--Auriga region is motivated by its close
proximity, the relative isolation in which its YSOs appear to form,
and the extensive literature on this region including the objects of
our sample (see references throughout this paper). By focusing on a
single star-forming region, the effects of different environment are
minimized. Systematic infrared and millimeter surveys of Taurus have
identified most of the embedded class~I objects in this cloud down to
1~mm fluxes of 10~mJy, corresponding to envelope masses of
0.0015~M$_{\sun}$ (Tamura et~al.\ \markcite{tam91}1991, Ohashi et~al.\
\markcite{oha96a}1996). As a subset of this complete sample, our nine
objects are well defined and representative of the embedded phase.

The outline of the paper is as follows. In \S 2 the nine sources
studied in this paper are introduced, together with their selection
from the sample defined by Tamura et~al.\ \markcite{tam91}(1991). The
observations are discussed in \S 3. In \S 4 the millimeter-continuum
emission of the disks and envelopes around the sources are analyzed.
The interferometer observations and literature values of the 1.1~mm
single-dish fluxes are used to obtain `pure' envelope fluxes. The
results of the HCO$^+$ observations are presented in \S 5, and further
analyzed together with the envelope fluxes within the framework of the
spherically symmetric inside-out collapse model of Shu
\markcite{shu77}(1977) and related power-law density distributions in
\S 6. The implications for the full sample of Tamura et~al.\
\markcite{tam91}(1991) are discussed in \S 7, where a relative
evolutionary ordering of the objects is proposed. Our main conclusions
are summarized in \S 8.

% -----------------------------------------------------------------

\section{Source sample}

Our sources have been chosen from the flux- and color-limited sample
of 24 IRAS sources located in Taurus--Auriga identified as embedded
YSOs by Tamura et~al.\ \markcite{tam91}(1991, hereafter TGWW). Their
main selection criteria are infrared color $\log \bigl[F_\nu
(25\,\mu{\rm m})/ F_\nu (60\,\mu{\rm m})\bigr] < -0.25$, and infrared
flux $F_\nu > 5$~Jy at either 60 or 100~$\mu$m. Sources with a known
identification as an SAO star or a galaxy were rejected. In Table~2
the sources of TGWW\markcite{tam91} are listed with some of their
properties. One source (04154+1755) is listed as a galaxy in the
SIMBAD database, and is dropped from the sample. The color criterion
limits the sources to objects more embedded than the majority of
T~Tauri stars; only eight out of 23 are optically visible T~Tauri-like
objects. The bolometric luminosity of the sample ranges between 0.7
and 22 $L_{\sun}$.  Near-infrared imaging of the sample by
TGWW\markcite{tam91} and Kenyon et~al.\ \markcite{ken93b}(1993b) shows
that most sources have associated reflection nebulosity with sizes
between 1000 and 3000~AU. Seven out of the 23 sources have a clear
monopolar or bipolar morphology, suggesting that the bipolar outflow
plays an important role in the appearance of YSOs at these short
wavelengths. CO outflow emission is detected toward 20 out of the 23
objects (Moriarty-Schieven et~al.\ \markcite{mor94}1994).

One source, Haro~6--10 (04263+2426; identified with GV~Tau = Elias
3--7, located in the L1524 cloud), was added to this sample following
Kenyon, Calvet \& Hartmann \markcite{ken93a}(1993a) and Leinert \&
Haas \markcite{lei89}(1989). The latter authors show that this object
is a T~Tauri star (GV~Tau) with a more embedded companion. Its
25--60~$\mu$m color index of $-0.19$ is only just above the selection
criterion of TGWW\markcite{tam91}.

\placefigure{f1}
\placetable{t1}
\placetable{t2}

Fourteen of the 24 objects of the sample were found to show HCO$^+$
3--2 emission of $T_{\rm mb} > 0.5$~K in the $19''$ beam of the James
Clerk Maxwell Telescope (see \S 3.2, Fig.~1, and Table~2).  The nine
strongest of these objects were subsequently selected for our
subsample, and are listed in Table~1 with some of their basic
properties.  Seven are embedded sources (class~I according to the
classification of Lada \markcite{lad87}1987); T~Tau and Haro~6--10
(GV~Tau) are optically visible (class~II), but have more embedded
companions. All nine show evidence of CO outflow emission (Terebey,
Vogel \& Myers \markcite{ter89}1989; Moriarty-Schieven et~al.\
\markcite{mor94}1994; Hogerheijde et~al.\ \markcite{mrh98a}1998a).  It
should be kept in mind that, given the observed binary frequency of
T~Tauri stars of 40--60\% (Ghez et~al.\ \markcite{ghe93}1993; Leinert
et al. \markcite{lei93}1993), four to six of our nine sources may in
fact be multiple systems. In addition to Haro~6--10 and T~Tau (Dyck,
Simon \& Zuckerman \markcite{dyc82}1982; Ghez et~al.\
\markcite{ghe93}1993), companions have been reported for L1551~IRS~5
(Looney et~al.\ \markcite{loo96}1996) and L1527~IRS (Fuller, Ladd \&
Hodapp \markcite{ful96}1996).

The pointing centers of our observations were based on optical or
near-infrared observations, which have an accuracy of
2$''$--3$''$. Because of its deeply embedded nature, no reliable
position of L1527~IRS was available, and the IRAS position was
used. In Table~1 the positions derived from the
millimeter-interferometer data are listed (see \S 4.1), which have an
accuracy of $\sim 1''$. Only for L1535~IRS and L1527~IRS do these
positions differ by 4$''$--5$''$ from the values listed by
TGWW\markcite{tam91}.

Upper limits to the mass of the central stars can be obtained by
assuming that all bolometric luminosity is stellar and that the object
is located on the birthline in the Hertzsprung--Russell diagram
(Stahler \markcite{sta88}1988; Palla \& Stahler \markcite{pal93}1993).
The inferred maximum masses range between 0.15~$M_\odot$ for TMC~1 and
2.6--2.7~$M_\odot$ for L1551~IRS~5 and T~Tau, and are listed in
Table~1. For T~Tau N and S masses of 2 and 1~$M_\odot$, respectively,
are inferred by Bertout \markcite{ber83}(1983) and Beckwith et~al.\
\markcite{bec90}(1990). Leinert \& Haas \markcite{lei89}(1989) quote a
mass of 1.0--1.5~$M_\odot$ for Haro~6--10. Both of these objects are
located close to the birthline, and the inferred values from Table~1
are near their true masses.

As part of the TGWW\markcite{tam91} sample, our sources have been
studied previously by various authors. Moriarty-Schieven et~al.\
(\markcite{mor92}1992, \markcite{mor94}1994, \markcite{mor95}1995)
observed the sample in single-dish continuum emission at 800~$\mu$m
and 1.1~mm, and in transitions of CO, CS and H$_2$CO, deriving
envelope masses and beam-averaged densities. We will use their 1.1~mm
fluxes in our analysis. Kenyon et~al.\ (\markcite{ken93a}1993a,
\markcite{ken93b}1993b) modeled the spectral energy distributions
(SEDs) of an overlapping sample of YSOs, and observed and modeled the
near-infrared emission. They found that the SEDs and the near-infrared
images can be reproduced by the collapse model of Terebey et~al.\
\markcite{ter84}(1984), but that a cleared-out bipolar cavity is
required in many cases to simultaneously fit the near-infrared and
far-infrared emission. Ohashi et~al.\ (\markcite{oha91}1991,
\markcite{oha96a}1996) performed interferometric observations of CS
2--1 of seven objects of our sample, as well as of some more evolved
T~Tauri objects.

% -----------------------------------------------------------------

\section{Observations}

\placetable{t3}

An overview of the data obtained for each source is given in Table~3.
In the following the details of the observations are discussed.

% ------------------------------

\subsection{Interferometer observations of millimeter-continuum emission}

Observations of the continuum emission at 3.4~mm and 2.7~mm were
obtained with the Owens Valley Radio Observatory (OVRO) Millimeter
Array%
\footnote{The Owens Valley Millimeter Array is operated by the
California Institute of Technology under funding from the U.S.\
National Science Foundation (\#AST93--14079).}
between 1992 and 1997, simultaneously with the HCO$^+$ 1--0, and the
$^{13}$CO and C$^{18}$O 1--0 transitions, respectively. During the
3.4~mm observations the array consisted of five antennas; the 2.7~mm
observations were made with a six-element array. Two sources were
observed per track. Data taken in the low-resolution and equatorial
configurations were combined, resulting in a $uv$~coverage with
spacings between 4 and 40 k$\lambda$ at 3.4~mm and between 4 and 80
k$\lambda$ at 2.7~mm. This corresponds to naturally weighted,
synthesized beams of $6''$ and $3''$~FWHM, respectively.  The
observations of T~Tau were made in five different array configurations
(van~Langevelde, van~Dishoeck \& Blake \markcite{hvl94a}1994a). The
lower and upper-sideband continuum signals were recorded separately
over the full instantaneous 1~GHz IF-bandwidth.  The data were
calibrated using the MMA package, developed specifically for OVRO
(Scoville et~al.\ \markcite{sco93}1993). The quasars PKS~0333+321 and
0528+134 served as phase calibrators (0420$-$014 for the observations
of T~Tau); the amplitudes were calibrated on 3C~454.3 and 3C~273,
whose fluxes at the time were determined from observations of the
planets.

The interferometer data were edited in the usual manner by flagging
data points with clearly deviating amplitudes and phases. Editing was
especially necessary for day-time observations at 2.7~mm, when the
phase stability of the atmosphere can be low. The final $1\sigma$ RMS
noise value of the visibilities is approximately 4~mJy when vector
averaged over 10~k$\lambda$ wide $uv$~intervals (see \S 4.1). The
naturally weighted, cleaned images have a typical $1\sigma$ noise
level of 2~mJy\,bm$^{-1}$. Reduction and analysis of the visibility
data was carried out within the MIRIAD software package.

% ------------------------------

\subsection{Single-dish observations of HCO$^+$ and H$^{13}$CO$^+$
emission}

Maps of HCO$^+$ $J=$~1--0 (89.18852~GHz) emission were obtained with
the IRAM~30m telescope, covering regions between $112''\times 112''$
and $168''\times 168''$ (approximately 16,000--24,000~AU in diameter;
see Fig.~4). The maps were sampled at intervals between $12''$ and
$28''$ depending on the source, with a beam size of $28''$. The data
were obtained in position-switched mode, with a typical switch of
15$'$--30$'$ in right~ascension. Special care was taken to ensure that
the off~positions were free of HCO$^+$ emission. Pointing was checked
regularly, and the maps were obtained in such a way as to minimize
systematic effects of pointing drifts.  The remaining pointing error
is smaller than $5''$.  The spectra were obtained at a frequency
resolution of 40~kHz (0.14~km\,s$^{-1}$), and were converted to the
main-beam temperature scale using $\eta_{\rm mb}=0.60$, resulting in
an RMS noise level of typically 0.4~K per channel.  For T~Tau a region
of $30''\times 30''$ was mapped with a grid spacing of $15''$, and a
RMS noise level of 0.16~K at a velocity resolution of
0.33~km\,s$^{-1}$.

In the HCO$^+$ $J=$~3--2 (267.55762~GHz) and 4--3 (356.73429~GHz)
transitions the sources were mapped over a $40''\times 40''$
($5600\times 5600$~AU) region at the James Clerk Maxwell Telescope%
\footnote{The James Clerk Maxwell Telescope is operated by the Joint
Astronomy Centre, on behalf of the Particle Physics and Astronomy
Research Council of the United Kingdom, the Netherlands Organization
for Scientific Research, and the National Research Council of
Canada.}%
 (JCMT). The maps were sampled at 5$''$--10$''$ intervals,
corresponding to ${1\over 2}$\,--\,${2\over 3}$ times the beam sizes
of $19''$ (267~GHz) and $14''$ (356~GHz), respectively. The
observations were obtained in position-switched mode, with a typical
switch of 15$'$--30$'$ in right~ascension, again to ensure that the
spectra are not contaminated by HCO$^+$ emission at the off
position. Pointing was checked regularly, and the remaining
uncertainty is less than $5''$.  The spectra were recorded with the
Digital Autocorrelation Spectrometer (DAS) backend with a typical
resolution of 156~kHz (0.15--0.18~km\,s$^{-1}$). Since the local
oscillator of the 267~GHz receiver at the JCMT has no phase-lock loop,
the HCO$^+$ 3--2 spectra have a minimum effective line width of
$\sim$~0.5--1.0~km\,s$^{-1}$. The spectra have been converted to the
main-beam antenna temperature scale using $\eta_{\rm mb}=0.69$
(267~GHz) and $\eta_{\rm mb}=0.58$ (356~GHz), obtained from
measurements of Jupiter and Mars by the JCMT staff.  Typical resulting
RMS noise levels are 0.2--0.4~K per channel on the main-beam
temperature scale.

Observations of HCO$^+$ 3--2 toward all 24 sources of
TGWW\markcite{tam91} were obtained with the JCMT in 1993 August.
During these observations the effective calibration was ill determined
because of technical difficulties. From comparison with spectra
obtained in 1995, main beam efficiencies between $\eta_{\rm mb} =
0.69$ and 0.31 were obtained, resulting in a calibration uncertainty
of at least 30\%.

Observations of H$^{13}$CO$^+$ 3--2 (260.25548~GHz) and 4--3
(346.99854~GHz) were also obtained at the JCMT toward the source
positions.  Instead of a position switch, a beam switch of $180''$ was
used in 1996 May. Typical RMS noise levels are 60--100~mK on the
main-beam temperature scale at 0.14~km\,s$^{-1}$ resolution.
Nobeyama~45m telescope maps in H$^{13}$CO$^+$ 1--0 of five of our
sources have been presented by Mizuno et~al.\
\markcite{miz94}(1994). The single-dish observations were further
reduced and analyzed with the CLASS software package.

% -----------------------------------------------------------------

\section{Disk and envelope continuum emission}
\subsection{Millimeter-continuum visibilities}

In order to study the evolution of the circumstellar disks and
envelopes around YSOs, it is necessary to accurately separate their
respective contributions to the millimeter-continuum emission. For
this purpose, the interferometer observations are best represented in
the $uv$~plane. Since visibilities are complex quantities, only
combinations of the real and imaginary parts can be plotted. The
vector average of the amplitudes with respect to a given phase center
corresponds to the Fourier transform of those components of the
sky-brightness distribution which are symmetric around that
position. It is important to use the correct source position as the
phase center, because an offset of the order of the synthesized beam
(3$''$--6$''$) or more creates an artificial decrease of flux with
$uv$~distance. Since amplitudes are never negative, random noise in
the complex visibilities translates to a non-zero expectation value
for the vector-averages even in the absence of emission.

\placefigure{f2}

In Fig.~2 the vector-averaged and $uv$-binned visibilities of our
sources are shown as functions of $uv$~distance, with their $1\sigma$
error bars and zero-signal expectation values. The positions of the
unresolved point sources were adopted as phase centers (see below and
Table~1). As a reference, the naturally weighted, cleaned images are
also presented.  All nine sources are detected in 3.4 and 2.7~mm
continuum emission, with total fluxes ranging between 4 and
200~mJy. Three sources, L1489~IRS, L1535~IRS and TMC~1, are detected
above the zero-signal expectation value in only one or two $uv$~bins.
However, the continuum positions listed in Table~1 agree well with the
2~$\mu$m positions, and the objects can be clearly discerned in the
cleaned images.  It is therefore concluded that these are true
detections. The other six sources are confidently detected above the
noise level in all $uv$~bins.

Because the vector-averaged fluxes form the Fourier transform of the
sky brightness, the signal of a point source is constant with
$uv$~distance in the absence of noise, while that of an extended
structure is a decreasing function of $uv$~separation. Emission on
scales larger than $\sim 50''$ (corresponding to the shortest spacing
of $\sim 4$~k$\lambda$) is resolved out altogether by the array; on
intermediate scales only part of the flux is recovered, depending on
the $uv$~coverage. Such resolved emission is detected on top of
unresolved emission toward five sources: T~Tau, Haro~6--10,
L1551~IRS~5, TMR~1 and L1527~IRS. The uncertain $uv$~dependence of the
visibilities of the weaker sources L1489~IRS, L1535~IRS, and TMC~1
precludes any statement about the spatial distribution of their
continuum emission.

\placetable{t4}

The observed visibilities, and not just their vector averages, were
fit with a source model consisting of a point source plus an extended
Gaussian component whenever necessary. This method has been applied
previously to the millimeter-continuum emission of YSOs by, e.g.,
Keene \& Masson \markcite{kee90}(1990), Terebey et~al.\
\markcite{ter93}(1993), and Butner et~al.\
\markcite{but94}(1994). These authors, however, use a complete
envelope model to describe the extended component. Because of the
limited signal-to-noise ratio and $uv$~coverage of our data, such
level of detail is unwarranted.  The obtained fit parameters
(position, point-source flux, flux and FWHM of the extended component)
are listed in Table~4. The unresolved point sources have sizes $<3''$
($<420$~AU in diameter). Their positions are listed in Table~1 and
agree within 2$''$ with the near-infrared positions, except for
L1527~IRS and L1535~IRS, which are shifted by $\sim 5''$ to the
northeast and north, respectively.  Ladd et~al.\
\markcite{ldd91}(1991) and Fuller et~al.\ \markcite{ful96}(1996) found
a similar shift for L1527~IRS from single-dish millimeter continuum
maps. In the following, only the point-source fluxes will be used
quantitatively in the analysis; the inferred values for the extended
sources are not sufficiently reliable for further interpretation.

Of the sources in our sample only Haro~6--10 had not been observed
previously by millimeter interferometry; all others have been observed
before, though generally at a lower signal-to-noise ratio, resulting
in fewer detections. The total 2.7~mm flux levels range between 4~mJy
and 200~mJy, and agree well with those listed by Keene \& Masson
\markcite{kee90}(1990), Ohashi et~al.\ (\markcite{oha91}1991,
\markcite{oha96a}1996) and Terebey et~al.\ \markcite{ter93}(1993). At
the position of the reported companion of L1527~IRS (Fuller et~al.\
\markcite{ful96}1996) no continuum emission is detected at 2.7~mm at
an estimated $1\sigma$ noise level of 4~mJy\,bm$^{-1}$, taking into
account primary beam attenuation and $uv$~coverage.

The unresolved point-source emission seen by millimeter
interferometers around YSOs is usually attributed to thermal emission
from an optically thick circumstellar disk (e.g., Terebey et~al.\
\markcite{ter93}1993).  Lay et~al.\ \markcite{lay94}(1994) and Mundy
et~al.\ \markcite{lgm96}(1996) have resolved the circumstellar disks
around HL~Tau and L1551~IRS~5 with the JCMT--CSO single-baseline and
BIMA interferometers. They find sizes of 60--80~AU for the semi-major
axes of these disks, consistent with the upper limits found here.
Spectral indices over $\lambda = 2.7$~mm to 870~$\mu$m of $\alpha =
2.7$ and 2.5, respectively, are inferred. The recent identification of
L1551~IRS~5 as a binary system instead of a resolved disk (Looney
et~al.\ 1997) does not influence the inferred value for
$\alpha$. Van~Langevelde et~al.\ \markcite{hvl95}(1995) find $\alpha
\approx 2.5$ between 2.7~mm and 840~$\mu$m for T~Tau, although
possible variability and decomposition into thermal and non-thermal
contributions for T~Tau N and S complicate the interpretation (see
also Hogerheijde et~al.\ \markcite{mrh98b}1998b).

Non-thermal contribution to the continuum flux of YSOs at millimeter
wavelengths could come from free--free emission of an ionized wind.
Fluxes at 2, 3.6 and 6~cm are given for some of our sources by
Rodr{\'\i}guez et~al.\ \markcite{rod89}(1989), Keene \& Masson
\markcite{kee90}(1990) and Skinner \& Brown
\markcite{ski94}(1994). From these values and the expected flat
free--free spectrum, it is estimated that this mechanism does not
provide more than 10\% of the 2.7~mm flux of the YSOs observed here
(cf.\ Keene \& Masson \markcite{kee90}1990; Lay et~al.\
\markcite{lay94}1994). Only for T~Tau may up to 30\% of the 2.7~mm
flux be attributed to free--free emission (Skinner \& Brown
\markcite{ski94}1994; van~Langevelde et~al.\ \markcite{hvl95}1995;
Hogerheijde et~al.\ \markcite{mrh98b}1998b).

The millimeter-continuum visibilities show that at least two-thirds of
the embedded sources selected here are surrounded by compact
disks. This detection rate is significantly larger than that of 15\%
reported by Ohashi et~al.\ \markcite{oha96a}(1996) for a largely
identical sample, owing entirely to the differences in sensitivity
(2~mJy\,bm$^{-1}$ vs.~3--7~mJy\,bm$^{-1}$). The inferred 2.7~mm disk
fluxes range between 7 and 100~mJy, which may correspond to ranges in
disk size, density, temperature, orientation, or mass. It is therefore
not straightforward on the basis of these fluxes alone to investigate
evolutionary effects, such as disk growth. Comparison with point
source fluxes of more evolved class~II objects is further complicated
by the fact that this phase lasts ten times longer than the embedded
phase. During this evolutionary period significant dispersal of the
disk is expected, and it is not clear which objects represent the
youngest class~II sources (cf.~Ohashi et~al.\
\markcite{oha96a}1996). Dutrey et~al.\ \markcite{dut96}(1996) find an
average flux at 2.7~mm of $\sim 24$~mJy for a sample of 12 T~Tauri
stars with the Plateau de Bure Interferometer, comparable to the flux
levels found toward our embedded sources.

% ------------------------------

\subsection{Separating envelope and disk flux at $\lambda = 1.1$~mm}

Moriarty-Schieven et~al.\ \markcite{mor94}(1994) obtained the
continuum flux at 1.1~mm of the TGWW\markcite{tam91} sample in a
$19''$ beam with the JCMT. The flux of Haro~6--10 is given by Kenyon
et~al.\ \markcite{ken93a}(1993a). These single-dish measurements
contain contributions from the extended envelope as well as any
compact disk. The point-source fluxes derived in the previous section
can be used to separate these components if a spectral index for the
point-source emission between 3.4/2.7~mm and 1.1~mm is known or
adopted. Since all our sources have good signal-to-noise single-dish
fluxes, reliable envelope fluxes can be obtained even for those
sources which have only marginal detections in the interferometer
beam.

For typical densities and values of the dust emissivity, the envelope
is optically thin at 1.1~mm, and its flux traces all the mass within
the beam. The unresolved circumstellar disks, on the contrary, are
likely to be optically thick. For a sharp-edged, isothermal,
unresolved disk and optically thick emission, a spectral slope of
$\alpha=2.0$ is expected. More realistic disks, with radial
surface-density and temperature gradients, may have $\alpha>2$,
because the optically thick area increases for shorter wavelengths. A
maximum value for $\alpha$ is found from the ratio of the 3.4/2.7~mm
OVRO and 1.1~mm JCMT fluxes, and ranges between 2.4 and 4.0. These are
strict upper limits since extended emission is actually observed
toward half of the sources in the interferometer beam. In the
following, $\alpha=2.5$ will be assumed as a best estimate for the
spectral slope of the disk based on the measured values for HL~Tau,
L1551~IRS~5, and T~Tau. The effect of over- or underestimating
$\alpha$ is to attribute too little, or too much, respectively, of the
1.1~mm flux to the envelope. Sources with relatively weak fluxes at
1.1~mm are worst affected by the uncertainty in $\alpha$. However,
none of the results from the subsequent analysis depend critically on
the assumed value.

\placetable{t5}

In Table~5 the 1.1~mm single dish fluxes from Moriarty-Schieven
et~al.\ \markcite{mor94}(1994) and the estimated envelope
contributions are listed. The envelopes and disks are found to
contribute equally to the single-dish fluxes: approximately 30--75\%
of the single-dish flux at 1 mm originates in the envelope; for
$\alpha=2.0$ this would be 50--85\%. Toward Haro~6--10 all single-dish
flux can be attributed to the point source detected at 2.7~mm,
although an extended component appears also present in the
visibilities (see Fig.~2). No constraints on the spatial distribution
of the 3.4 and 2.7~mm emission were obtained in \S 4.1 for L1489~IRS,
L1535~IRS and TMC~1, resulting in a range of possible envelope
fluxes. These caveats are included in the estimated error bars on the
fluxes in the analysis below.

% ------------------------------

\subsection{Mass estimates for the envelopes and disks}

The envelope mass traced in the $19''$ JCMT beam can be obtained,
assuming optically thin emission from the relation
\begin{equation}
M = {{F_\nu D^2}\over{\kappa_\nu \, B_\nu(T_{\rm d})}} \; \*
\biggl({{\tau_\nu}\over{1-e^{-\tau_\nu}}}\biggr),
\end{equation}
where $M$ is the mass, $F_\nu$ is the 1.1~mm (envelope) flux, $D$ is
the distance to Taurus (140~pc), $\kappa_\nu$ is the dust emissivity
at 1.1~mm per unit total mass, and $B_\nu (T_{\rm d})$ is the Planck
function at the dust temperature $T_{\rm d}$. For $\kappa_\nu$ a value
of 0.01~cm$^2$\,g$^{-1}$ is adopted (Agladze et~al.\
\markcite{agl94}1994; Pollack et~al.\ \markcite{pol94}1994; Ossenkopf
\& Henning \markcite{oss94}1994). A dust temperature $T_{\rm d} =
30$~K is assumed, consistent with the average temperature inferred
from SED fitting by Moriarty-Schieven et~al.\
\markcite{mor94}(1994). The obtained envelope masses range from
$<0.0014$~$M_\odot$ for Haro~6--10 to 0.26~$M_\odot$ for L1551~IRS~5,
and are listed in Table~5.

Using the same expression, a lower limit to the mass of the disks can
be obtained, again assuming $\tau_\nu \ll 1$. Since the disks are
likely to be optically thick at 3.4/2.7~mm, only lower limits are
found. Using the same dust emissivity of 0.01~cm$^2$\,g$^{-1}$ scaled
by $(\lambda/1.1\,{\rm mm})^{-1.5}$ and a dust temperature of 30~K,
disk masses between $4\times 10^{-3}$~$M_\odot$ for L1489~IRS and
$7\times 10^{-2}$~$M_\odot$ for L1551~IRS~5 are inferred (see
Table~4), i.e., a factor of $\sim 3$ less than the envelope
masses. For a higher dust temperature the inferred disk masses
decrease as $M \propto T_{\rm d}^{-1}$, while for significant optical
depth they increase as $M \propto \tau_\nu$ for $\tau \gg 1$. Given
the fact that the disks likely have $\tau_\nu \gtrsim 1$, it is
concluded that the disks and envelopes typically are equally massive,
with each carrying $\sim 10\%$ of the mass of the central object
(cf.~Table~1).

% ------------------------------

\section{HCO$^+$ and H$^{13}$CO$^+$ emission}

\placefigure{f3}
\placefigure{f4}
\placetable{t6}

In Fig.~3 the HCO$^+$ 1--0, 3--2 and 4--3, and H$^{13}$CO$^+$ 3--2 and
4--3 spectra obtained toward the source positions are presented. In
Fig.~4 contour maps of the emission integrated over the full width of
the HCO$^+$ lines are shown. The 50\% intensity contours are indicated
by the thick solid lines in all maps. The observed velocity-integrated
intensities toward the source centers and the FWHM of the maps are
listed in Table~6.

Toward the majority of sources the HCO$^+$ spectra are dominated by a
relatively narrow line ($\Delta V\approx$
2--3~km\,s$^{-1}$). Double--peaked line profiles are present toward
most sources. Since care was taken to obtain emission-free reference
positions (\S 3.2), these can be attributed to self-absorption.
Red-shifted self-absorption features in a high-excitation line like
HCO$^+$ 4--3 are sometimes invoked as a tracer of protostellar infall
(Walker et~al.\ \markcite{wal86}1986; Zhou et~al.\
\markcite{zho93}1993; Zhou \markcite{zho95}1995; Ward-Thompson
et~al.\ \markcite{war96}1996). It should be noted that such features
are only prominently present toward two of our sources, L1489~IRS and
L1527~IRS.  The occurrence of blue-shifted absorption, or the absence
of any features altogether, indicates that other factors like
orientation are equally important in determining the shape of the
line profiles. For example, toward T~Tau the absorption seen in
interferometer data is filled in by emission from the surrounding
cloud in the single-dish observations (van~Langevelde et~al.\
\markcite{hvl94a}1994a).

In many spectra, emission from the bipolar outflow is visible as line
wings, but only toward T~Tau do the wings dominate the total
intensity. The integrated intensities listed in Table~6 and the
contour maps shown in Fig.~4 therefore predominantly reflect the
quiescent envelope material. For T~Tau, estimates of the relative
contributions of the envelope and the outflow are given in the
table. The H$^{13}$CO$^+$ lines are narrow, with $\Delta V
=0.5$--3.0~km\,s$^{-1}$, peaking at the same velocity as the HCO$^+$
absorption features, if present.

The average optical depth $\bar\tau$ over the HCO$^+$ line profiles
can be estimated from the ratio of the HCO$^+$ and H$^{13}$CO$^+$
integrated intensities, assuming an abundance ratio of 65:1 for
[HCO$^+$]\,:\,[H$^{13}$CO$^+$]. Typical opacities of 4--10 or less are
found (see Table~6). Larger optical depths are inferred for the 1--0
lines of L1527~IRS and TMC~1A, with $\bar\tau \approx $~18--21. The
H$^{13}$CO$^+$ lines are optically thin in all observed transitions
and toward all sources.

A beam-averaged HCO$^+$ abundance can be obtained from the
H$^{13}$CO$^+$ 1--0 intensities toward L1551~IRS~5, L1535~IRS, TMR~1,
and L1527~IRS (Mizuno et~al.\ \markcite{miz94}1994), and the C$^{18}$O
1--0 data of Hayashi et~al.\ \markcite{hay94}(1994), in respective
beams of $19''$ and $15''$. Assuming LTE at an excitation temperature
$T_{\rm ex}$, the beam-averaged column density in cm$^{-2}$ is given
in cgs~units by
\begin{equation}
\bar N = 10^5\times {{3k^2\over{4h\pi^3\mu^2\nu^2}}} \; \*\allowbreak
\exp \biggl( {{h\nu J_l}\over{2k T_{\rm ex}}} \biggr) \; \*\allowbreak
{{T_{\rm ex} + h\nu \,/\, 6k(J_l + 1)} \over {e^{-h\nu / kT_{\rm ex}}}} 
\;\*\allowbreak
\int T_{\rm mb} \,\biggl({{\tau}\over{1-e^{-\tau}}}\biggr)\, dV
\end{equation}
(Scoville et~al.\ \markcite{sco86}1986), where $\int T_{\rm mb} dV$ is
the integrated intensity in K\,km\,s$^{-1}$ of the $J_u$--$J_l$
transition with frequency $\nu$ and opacity $\tau$. The permanent
dipole $\mu$ is 0.112~Debye and 3.91~Debye for C$^{18}$O and
H$^{13}$CO$^+$, respectively (Millar et~al.\ \markcite{mil91}1991
). With critical densities of $2\times 10^5$~cm$^{-3}$ and $2\times
10^3$~cm$^{-3}$, respectively, the excitation of H$^{13}$CO$^+$ 1--0
and C$^{18}$O 1--0 is likely to be thermalized, especially in the
inner dense regions of the envelope. Assuming the same abundance ratio
for [HCO$^+$]\,:\,[H$^{13}$CO$^+$] of 65:1, a ratio of
[H$_2$]\,:\,[C$^{18}$O] of $5\times 10^6$, and an excitation
temperature of 30~K, an abundance of HCO$^+$ of $(1.2 \pm 0.4)\times
10^{-8}$ is found. The uncertainty in this value is dominated by the
spread in the data points of the five sources. This value is
comparable to that inferred for dark clouds of $\sim 8\times 10^{-9}$
(Irvine et~al.\ \markcite{irv87}1987), and it is assumed that it can
be applied to all sources in our sample.

In the maps of integrated HCO$^+$ 1--0 intensity, each source shows up as
a distinct core of roughly 50$''$--150$''$ in diameter ($28''$ beam
size), superposed on extended emission from the surrounding cloud. The
enhancement in integrated intensity around the sources arises
partially from the increased line widths. At positions $\sim 60''$
(8400~AU) away from the center, the average HCO$^+$ 1--0 line width
has decreased from 2.0--3.0~km\,s$^{-1}$ to 0.5--1.0~km\,s$^{-1}$. The
emission in HCO$^+$ 3--2 and 4--3 is much more concentrated around the
sources, with typical sizes of 20$''$--30$''$ in diameter compared
with the respective beam sizes of $19''$ and $14''$.  The 50\%
intensity contours are nearly circular and all cores appear marginally
resolved with FWHM $\sim 1.5$ times the beam size.  This is typical
for density distributions following a radial power law, i.e.,
distributions without an intrinsic size scale within the sampled
range.  This point is illustrated by Ladd et~al.\
\markcite{ldd91}(1991) and Terebey et~al.\ \markcite{ter93}(1993) for
continuum emission; line emission does possess an intrinsic density
scale. The HCO$^+$ 3--2 and 4--3 emission around T~Tau shows a
prominent extension toward the southwest, which is also seen in
several transitions of CO (Edwards \& Snell \markcite{edw82}1982;
Schuster et~al.\ \markcite{sch93}1993) and is probably associated with
the reflection nebula NGC~1555.

For the scope of this paper, i.e., the comparison of the continuum
flux with the HCO$^+$ line emission as tracers of the circumstellar
envelopes, the spherically averaged quantities listed in Table~6
suffice. In a future paper (Hogerheijde et~al.\ \markcite{mrh98c}1998c),
the line profiles will be studied in greater detail.

% -----------------------------------------------------------------

\section{Analysis}

The observed integrated HCO$^+$ 3--2 and 4--3 intensities are compared
with the 1.1~mm envelope fluxes in Fig.~5, and are seen to correlate
well. Since the HCO$^+$ line intensities depend primarily on density
and mass, whereas the continuum fluxes scale with dust temperature and
mass, the apparent correlation suggests that both quantities primarily
trace envelope mass.

The inferred envelope continuum fluxes at 1.1~mm and the HCO$^+$
observations presented above allow further investigation into the mass
and density structure of the circumstellar material around the objects
of our sample. From the beam-convolved HCO$^+$ source sizes it was
concluded that the envelopes do not posses any intrinsic scale within
the sampled range, i.e., their density follows a radial power
law. Many models of protostellar collapse predict such a density
distribution. The precise value of the power-law index and its
evolution with time depend on the formulation of the problem and
distinguish the various proposed models (e.g., Shu
\markcite{shu77}1977, Terebey, Shu \& Cassen \markcite{ter84}1984,
Galli \& Shu \markcite{gal93a}1993, Fiedler \& Mouschovias
\markcite{fie92}1992, \markcite{fie93}1993, Boss \markcite{bos93}1993,
and Foster \& Chevalier \markcite{fos93}1993) . In this section the
envelope fluxes and integrated HCO$^+$ intensities will be compared
with one such model, namely the self-similar, inside-out collapse
model of Shu~(1977).  Although this model is obviously oversimplified,
it has the advantage that it depends on only two parameters: the sound
speed $a$, and the initial mass of the cloud core, or, equivalently,
its initial outer radius $R$. Since the objects in our sample are all
located within the similar environment of a single star-forming
region, the dependence on these parameters is minimized. Because of
the simple formulation of this model, it has been used widely for
comparison to observations (Zhou \markcite{zho92}1992; Zhou et~al.\
\markcite{zho93}1993; Choi et~al.\ \markcite{cho95}1995; Ceccarelli
et~al.\ \markcite{cec96}1996). Formally, the assumption of
self-similarity breaks down when the so-called collapse expansion wave
reaches the outer radius. However, we will continue to use
this formulation even after this time as a qualitative description.

% ------------------------------

\subsection{Parameters of the model calculations}

The inside-out collapse model has only two main parameters: the sound
speed $a$ and the initial core radius $R$. In addition, the dust
emissivity $\kappa_\nu$ at 1.1~mm (\S 4.3), the HCO$^+$ abundance (\S
5), and the temperature of the gas and the dust are required to
calculate the emergent continuum flux and HCO$^+$ line strengths. The
density configuration of the initial state is given by $\rho(r) = (a^2
/ 2\pi G)\, r^{-2}$, and its mass by $M(R) = 2 R a^2 / G$.  Estimates
for the sound speed in the Taurus cloud cores vary between $a=0.2$ and
0.43~km\,s$^{-1}$ (Terebey et~al.\ \markcite{ter84}1984). A value of
$a=0.3$~km\,s$^{-1}$ is used here, corresponding to the isothermal
sound speed for the typical temperatures of $\sim$~30--40~K derived by
Moriarty-Schieven et~al.\ \markcite{mor94}(1994). In addition, it is
the smallest value which can explain the observed 1.1~mm continuum
flux of L1551~IRS~5 ($\sim$~3~Jy). It should be noted that the
density, flux and mass are strongly dependent on $a$, with $\rho
\propto a^2$ and $M \propto a^3$.

Estimates for the core radius can be obtained from the HCO$^+$ 1--0
and 3--2 maps, and from the C$^{18}$O 2--1 emission observed at
positions more than 30$''$--60$''$ away from the source centers
(Hogerheijde et~al.\ \markcite{mrh98a}1998a). The density at these
positions has dropped to (0.3--3.0)$\times 10^4$~cm$^{-3}$,
representative of the surrounding cloud. With $a=0.3$~km\,s$^{-1}$,
this corresponds to a radius of $R=9000$--28,000~AU where the core
merges into the surrounding cloud. Since most of the continuum and
line emission originates from the central regions of the core,
especially for the HCO$^+$ 3--2 and 4--3 lines which possess critical
densities $\ge 10^6$~cm$^{-3}$, $R=9000$~AU ($65''$) will be used.
Another estimate of the core size can be obtained from the observed
line widths in the surrounding cloud, typically $\Delta V \approx
0.4$~km\,s$^{-1}$. A turnover radius can be defined where the kinetic
energy contained in these motions, $\sim {1\over 2} (\Delta V)^2$,
balances the gravitational binding to a $M_\star \approx 0.5$
$M_{\sun}$ object, $GM_\star/R$. From such an analysis, a value of $R
\approx 10,000$~AU is found, in good agreement with the adopted
$R=9000$~AU.

The total envelope mass depends strongly on $R$, with $M_{\rm env}
\propto R$, and is therefore not well constrained by the observations.
For $R=9000$ AU, $M_{\rm env} = 1.8$ $M_{\sun}$ is obtained, a factor
of 3 higher than the average stellar mass in Taurus. However, the
bipolar outflow may disperse part of the envelope, or even reverse
infall before all envelope mass has accreted onto the protostar.  The
mass-accretion rate corresponding to $a=0.3$~km\,s$^{-1}$ is $\dot
M_{\rm acc} = 6 \times 10^{-6}$~$M_{\sun}$\,yr$^{-1}$. If the
accretion rate onto the star is the same as this large scale envelope
accretion rate, a luminosity of $L_{\rm acc} = G M_\star \dot M_{\rm
acc} / R_\star = 32$~$L_{\sun}$ is generated, assuming a stellar mass
of 0.5~$M_{\sun}$ and radius of 3~$R_{\sun}$. This accretion
luminosity is much larger than the typical observed values for our
sources, but a non-constant accretion rate onto the star can
result in a significantly reduced luminosity during large periods of
time separated by short outbursts of high luminosity (cf.~FU~Orionis
events; see Hartmann, Kenyon \& Hartigan \markcite{har93}1993, and \S
6).

The inside-out collapse model is based on isothermal initial
conditions.  After the onset of collapse and the formation of a
central object, this assumption no longer holds. A simple power-law
behavior for the dust temperature distribution is assumed, $T_{\rm d}
\propto r^{-0.4}$, as is valid for any centrally heated envelope which
is optically thin to the photons carrying the bulk of the heating
(e.g., Rowan-Robinson \markcite{row80}1980). At a characteristic
radius of 1000~AU, $T_{\rm d}=30$~K is assumed, consistent with recent
calculations of the dust temperature distribution in YSO envelopes
(e.g., Ceccarelli et~al.\ \markcite{cec96}1996). Although the gas and
the dust are not closely coupled thermally at the lower densities, the
simplifying assumption $T_{\rm kin} = T_{\rm d}$ is made for all
radii. Recent calculations by Ceccarelli et~al.\
\markcite{cec96}(1996) indicate that the gas kinetic temperature does
not deviate from the dust temperature by more than a factor of $\sim
2$.

For comparison, calculations are also made for power-law density
distributions, $\rho(r) = \rho_0 \, (r/r_0)^{-p}$, with $p=1$, 2 and
3. The same outer radius $R$, dust emissivity, HCO$^+$ abundance and
temperature structure as for the collapse model are used.

% ------------------------------

\subsection{Comparison of model calculations with observations}

\placefigure{f5}

Since the continuum emission at 1.1~mm is optically thin, the flux as
observed in a $19''$ beam can be easily calculated as a function of
time by direct integration along all lines of sight. The results are
shown in Fig.~6a. It should be stressed that the time-axis has no
quantitative meaning, since the mass-accretion rate, and hence the
mass of the envelope as function of time, is strongly dependent on the
ill-constrained sound speed. Furthermore, the evolution is followed
beyond the point where the assumption of self-similarity breaks down,
as indicated by the vertical dashed line, in that the envelope fluxes
observed toward almost the entire source sample fall beyond this
line. The results are therefore more realistically presented as a
function of density at the arbitrary radius of 1000~AU. This
approach is used in Fig.~6b, which presents curves for the collapse
and three power-law models.

In Table~5 the H$_2$ number densities at 1000~AU corresponding to the
inferred envelope fluxes are listed for the collapse model curve;
values range between $1\times 10^4$ and $2\times 10^6$~cm$^{-3}$, and
depend on the adopted dust temperature. Since the disks are estimated
to contribute 30--75\% to the total 1.1~mm flux, the inferred
densities do not depend critically on the assumed value for the
spectral index $\alpha$. For Haro~6--10, where all single-dish flux
can be attributed to a disk, only an upper limit to the density is
found.

To calculate the HCO$^+$ emission of the model envelopes, the
excitation must to be solved together with the radiative transfer,
since the lines are generally optically thick. A one-dimensional Monte
Carlo code has been used, based on the description of Bernes
\markcite{ber79}(1979). The radiation field at all transition
frequencies is represented simultaneously by 200 model photons
propagating through the model envelope. The HCO$^+\!$--H$_2$ collision
rates of Monteiro \markcite{mon85}(1985) and Green
\markcite{gre75}(1975) have been used, together with the line
frequencies of Blake et~al.\ \markcite{gab87}(1987).  The model is
divided into 15 concentric shells, each with a density following the
inside-out collapse model or any of the three power-law descriptions,
a kinetic temperature, and fractional abundances for HCO$^+$ and
H$^{13}$CO$^+$. No abundance variations with radius have been adopted.
Instead of a detailed velocity field, a turbulent line width of
1.0~km\,s$^{-1}$ FWHM has been used throughout the model
envelope. This produces line widths comparable to the observed
spectra.  For the excitation of lines for which the envelope is
transparent (e.g., H$^{13}$CO$^+$), the details of the velocity field
are not important.  Significant optical depths are obtained for
HCO$^+$, but it is found that the resulting integrated line
intensities do not depend strongly on the details of the velocity
field, and stay well within the estimated uncertainty of a factor of
$\sim 2$ originating from the other parameters of the source model,
i.e., sound speed, temperature, and abundance.  After the molecular
excitation has converged, the radiative transfer is solved. The
resulting sky-brightness distribution is then convolved with the
appropriate beam sizes.

Line intensities and FWHM source sizes predicted by the Monte Carlo
code are shown in Fig.~6c and~6d as functions of density for the
four models. The behavior of the curves reflects the increasing
molecular column density with larger $n_{\rm H_2}$, resulting in
higher line intensities and larger source sizes. At high column
density, the intensities level off because of optical depth. Since the
curves are parameterized by the density at 1000~AU, the central
density is lower for the $p=1$ model than for the $p=3$ model at the
same $n_{\rm H_2}$. This explains why the model curves can cross. The
behavior of the collapse-model curve is best understood in terms of
time, indicated along the top of the panels, rather than density. For
example the decrease of the HCO$^+$ line intensities with time of
20--30~K\,km\,s$^{-1}$ to a few K\,km\,s$^{-1}$ between $t\sim
10^4$~yr and a ${\rm few}\times 10^6$~yr reflects the decreasing
density and column density of the envelope. The H$^{13}$CO$^+$ 1--0,
3--2 and 4--3 line intensities decrease from 10--15~K\,km\,s$^{-1}$ to
$<0.2$~K\,km\,s$^{-1}$ over the same time span. Lower bounds to the
observational sensitivities are typically 0.2--0.7~K\,km\,s$^{-1}$, so
that the data can probe the full parameter range.  The FWHM sizes of
the beam-convolved model HCO$^+$ emission vary from 70$''$--110$''$
(1--0), 30$''$--110$''$ (3--2), and 20$''$--60$''$ (4--3). The maximum
in the convolved source size around $t\sim 10^5$~yr corresponds to the
moment when the collapse expansion wave reaches the outer radius of
the envelope, and the density follows $\rho \propto r^{-1}$ over a
large region. A flatter density distribution results in a more
extended sky-brightness distribution, and hence a larger convolved
source size. The time variation of the column density and the
steepening of the density distribution drive the evolution of the
integrated intensities and FWHM source sizes. The small observed
source sizes in 3--2 and 4--3 exclude density distribution as flat as
$p<1$. Otherwise, no strong constraints are placed on the model
parameters by the source sizes.

\placefigure{f6}

To investigate how well the different models can describe the
envelopes of our sources, the calculated HCO$^+$ intensities are
plotted against the corresponding envelope fluxes in Fig.~5. The
individual sources are identified in the HCO$^+$ 3--2 panel. Whereas
the continuum flux only traces mass, the HCO$^+$ lines probe both
density and mass.  For the inside-out collapse model, time runs from
the upper-right to lower-left corners of the panels; the density at
1000~AU decreases in that same direction for all four curves. The
observations are plotted with error bars corresponding to the $\sim
20$\% calibration uncertainty. For T~Tau, the relatively large
contribution made by the outflow to the HCO$^+$ 4--3 line wings is
incorporated in the error bar. Overall, the model curves are seen to
describe the observations well. The uncertainty in the model
parameters and the scatter in the data points preclude any constraints
on the density model, although the H$^{13}$CO$^+$ 4--3 data seem to
favor a slope $p \le 2$. To a large extent, the agreement between
models and observations reflects the unsurprising fact that less
massive envelopes are weaker in both continuum and line
emission. However, the correlation between the envelope flux and the
HCO$^+$ intensity indicates that the observed range in either of these
quantities is not due to variations in the temperature of the bulk of
the dust or of the HCO$^+$ abundance. This means that within the $\sim
20''$ beam, no significant depletion of molecules has occurred. Only
for L1551~IRS~5 do the observations fall significantly below the model
curves in many of the HCO$^+$ and H$^{13}$CO$^+$ panels. One possible
explanation is offered by the identification of this source as
undergoing a FU~Orionis outburst (cf.~Mundt et~al.\
\markcite{mun85}1985), which may result in an enhanced dust
temperature and a larger continuum flux. For the sources in general, the
correspondence between model curves and observations is better for the
HCO$^+$ 3--2 and 4--3 lines, consistent with the maps, which show that
significant contribution to the 1--0 line comes from the surrounding
cloud. It is concluded that the HCO$^+$ 3--2 and 4--3 emission
is a good tracer of the envelope mass.
 
% -----------------------------------------------------------------

\section{Discussion}

In the previous section it was shown that the envelopes around our
nine sources, as traced by their 1.1~mm continuum emission and HCO$^+$
3--2 and 4--3 lines, are characterized by a radial power-law density
distribution with slopes $p=$~1--2. This is consistent with the
inside-out collapse model of Shu \markcite{shu77}(1977), but more
detailed comparisons with line profiles should be made before further
statements about the validity of any specific collapse model can be
made (see Hogerheijde et~al.\ \markcite{mrh98c}1998c for such an
analysis of the HCO$^+$ data presented here).  Especially the HCO$^+$
3--2 and 4--3 lines are found to be especially robust tracers of the
envelopes. Similar conclusions were reached by Blake et~al.\
\markcite{gab94}(1994) for NGC~1333~IRAS~4 and van~Dishoeck et~al.\
\markcite{evd95}(1995) for IRAS~16293$-$2422. A beam-averaged HCO$^+$
abundance of $(1.2\pm 0.4) \times 10^{-8}$~cm$^{-3}$ is inferred from
C$^{18}$O 1--0 observations, and is found to be well-matched to the
continuum fluxes, i.e., essentially undepleted abundances over $20''$
scales. Typical masses within the $19''$ JCMT beam of
0.001--0.26~$M_\odot$ are found.

\placefigure{f7}

Moriarty-Schieven et~al.\ \markcite{mor95}(1995) used observations of
CS 3--2, 5--4 and 7--6 to investigate the envelopes around all sources
from the TGWW\markcite{tam91} sample. These transitions probe
densities between $10^6$ and $10^7$~cm$^{-3}$ and temperatures between
14 and 66~K, and as such should be well suited as envelope tracers. In
Fig.~7 a comparison is made between their CS data and our HCO$^+$ 3--2
measurements for the full TGWW\markcite{tam91} sample. The correlation
of the HCO$^+$ 3--2 ($19''$ beam) with the CS 3--2 line ($43''$ beam)
and with the 5--4 line ($32''$ CSO beam) is found to be poor,
reflecting the large contribution by the surrounding cloud to these
transitions. For the 5--4 line ($20''$ JCMT beam) and the 7--6 line
($20''$ beam), essentially the same trend is observed as in HCO$^+$
3--2, although the HCO$^+$ 3--2 is, on average, stronger by a factor
of 5--10. Thus, HCO$^+$ may be a more sensitive and easier to observe
envelope tracer than CS.

\placefigure{f8}

Does the observed trend of HCO$^+$ and 1.1~mm continuum flux as
tracers of the envelopes also hold for the full sample defined by
TGWW\markcite{tam91}?  In particular, are those sources which remained
undetected in HCO$^+$ 3--2 also different in 1.1~mm dust continuum
from the selected sources? In Fig.~8 the HCO$^+$ 3--2 observations of
the full TGWW\markcite{tam91} sample are plotted against their 1.1~mm
fluxes from Moriarty-Schieven et~al.\ \markcite{mor94}(1994). This
plot is an extension of the HCO$^+$ 3--2 panel of Fig.~5, with the
only difference that no correction is made for the contribution of
possible disks to the continuum flux, which is not known for the other
sources. As a reference, the model curve for inside-out collapse is
also shown. Note that this plot shows that the point-source
subtraction is not critical for the conclusions of \S 6; the data
points of our source sample (indicated by filled symbols) are still
reasonably well fit by the model curve.

The sources which were not included in our sample are indicated by the
open symbols, and can be split into two groups: those with detected,
but weak HCO$^+$ 3--2 emission, and those undetected in this line. The
first group forms the low-brightness tail of the
HCO$^+$--\,1.1~mm-continuum distribution of our sample. These are
sources with envelopes of low mass, either because they already have
accreted much of their core material, or because they are
intrinsically low-mass objects. The YSOs in the second group are found
to span the same range in 1.1~mm flux as our subsample, and are also
weak in C$^{18}$O 3--2 (Hogerheijde et~al.\ \markcite{mrh98a}1998a),
indicating that the lack of HCO$^+$ 3--2 emission is due to a low
total column density, and not to a low HCO$^+$ abundance or low
density. It is therefore concluded that the 1.1~mm continuum flux of
these sources originates primarily in a circumstellar disk, much
smaller than the $19''$ JCMT beam, instead of an extended
envelope. Interferometric observations of millimeter continuum toward
several of these sources (DG~Tau: Dutrey et~al.\ \markcite{dut96}1996;
GG~Tau: Ohashi et~al.\ \markcite{oha96a}1996) are consistent with this
interpretation, assuming a spectral index of $\alpha=2.5$.  No
detectable HCO$^+$ 3--2 emission is expected from disks with such
small beam-filling factors at the noise level achieved by our
observations. Dutrey et~al.\ \markcite{dut97}(1997) detected an
integrated HCO$^+$ 3--2 intensity of 1.5 K\,km\,s$^{-1}$ toward GG~Tau
in the $9''$ IRAM~30m beam, consistent with our upper limit of $\sim
0.3$ K\,km\,s$^{-1}$ in the $19''$ JCMT beam for an unresolved source.

In the simple evolutionary picture sketched in \S 1, sources without
envelopes but with disks are more evolved than those still surrounded
by envelopes. Obviously, intrinsically less massive objects appear to
be more evolved if judged by the absolute mass of their envelope
alone. Bontemps et al. \markcite{bon96}(1996) and Saraceno et~al.\
\markcite{sar96}(1996) used the envelope mass, as traced by the 1.1~mm
flux, together with the bolometric luminosity to obtain an
evolutionary ordering of an extended sample of YSOs. If the mass
accretion rate $\dot M_{\rm acc}$ is assumed to be constant, then the
mass of the envelope at time $t$ is given by $M_{\rm env} = M_0 - \dot
M_{\rm acc}\, t$, where $M_0$ is the initial core mass. The mass of
the central object is given by $M_\star = \dot M_{\rm acc}\, t$.  The
bolometric luminosity $L_{\rm bol}$ is a measure of $M_\star$, whether
it is mainly stellar in origin, or whether it is assumed to be
generated by mass accretion onto the star, $L_{\rm bol} = L_{\rm acc}
= G M_\star \dot M_{\rm acc}/R_\star$, with $R_\star$ the radius of
the protostar.  In this simple scheme, the ratio of $M_{\rm env}$ over
$L_{\rm bol}$ gives the relative evolutionary `phase' of the object; a
YSO starts off with large $M_{\rm env}$ and low luminosity (low
$M_\star$), and evolves toward high luminosity (large $M_\star$) and
low $M_{\rm env}$. In Fig.~2 of Saraceno et~al.\
\markcite{sar96}(1996) this ratio corresponds to the relative distance
traveled along the evolutionary tracks.

\placefigure{f9}

Based on the results of the previous section, the HCO$^+$ 3--2 line
strength can be used to trace the envelope mass uncontaminated by any
contribution from a disk, which may be a problem with the 1.1~mm
flux. In Fig.~9, the cumulative distribution of the quantity $\int
T_{\rm mb} dV ({\rm HCO^+}\,3\!-\!2) \, / \, L_{\rm bol}$ is shown for
our nine sources, the five sources from the TGWW\markcite{tam91}
sample with weak but detected HCO$^+$, and the sources undetected in
HCO$^+$. For this last group, the $2\sigma$ upper limits are used as
if they were detections. The first two groups appear to follow a
similar distribution, clearly separated from the third. A
Kolmogorov--Smirnov test indicates that there is a chance of $<0.2\%$
that the third group is similar to either our sample or the union of
the weak-HCO$^+$ sources with our sample. This difference is based on
the strict upper limits acquired on their HCO$^+$ 3--2 line strengths;
it also shows, however, that they do not have the low bolometric
luminosity expected from young, but intrinsically low-mass
objects. It can therefore be concluded that these sources form a group
of more evolved objects than the YSOs selected in \S 2. The HCO$^+$
line strength seems to be a better signpost of young age than either
the 1.1~mm continuum flux or the IRAS color.

The assumption of a constant mass-accretion rate is not critical for
this conclusion. Two types of variations may exist in $\dot M_{\rm
acc}$. Firstly, the accretion rate may gradually change over time, and
probably decrease (cf.~Foster \& Chevalier \markcite{fos93}1993). This
results in an enhanced luminosity for young sources and a smaller
range in $M_{\rm env}/L_{\rm bol}$. Since a clear range in $M_{\rm
env}/L_{\rm bol}$ is observed, its use as an evolutionary marker is
not precluded by this type monotonic of change in $\dot M_{\rm
acc}$. Secondly, long periods of relatively low accretion rates may be
separated by short bursts of high accretion (FU~Orionis effect,
cf.~Hartmann et~al.\ \markcite{har93}1993).  Protostars are expected
to spend $\sim 10\%$ of their lifetime in a FU~Ori phase, which means
that only two or three sources of the TGWW\markcite{tam91} sample may
be in this phase, and only one of our nine sources. Indeed,
L1551~IRS~5 is sometimes referred to as undergoing a FU~Ori outburst
(Mundt et~al.\ \markcite{mun85}1985).  The effect of the sudden
increase in $L_{\rm bol}$ is that an object is shifted temporarily to
lower $M_{\rm env}/L_{\rm bol}$; i.e., it appears more evolved and
more massive than it really is. The short period spend in a FU~Ori
phase, as inferred from the very small number of known FU~Ori objects,
ensures that for the 24~objects studied here the statistical
distribution of $M_{\rm env}/L_{\rm bol}$ is not seriously affected by
an individual object undergoing a FU~Ori outburst.

A difficulty in using $\int T_{\rm mb} dV ({\rm HCO^+}\,3\!-\!2) \,/\,
L_{\rm bol}$ to trace evolution is the opacity of the HCO$^+$ 3--2 line.
In Fig.~9 the cumulative distribution of our sources is shown after
taking into account the known 3--2 line opacity (see Table~6). This
correction is small for most sources. In addition, it makes the
difference between sources detected and undetected in HCO$^+$ 3--2
larger, strengthening the conclusion that they form two distinct
evolutionary groups.

Along the $\tau$-corrected curve the position of our objects is
indicated, giving a relative evolutionary ordering of our sample. Note
that no absolute timescale can be assigned, and that the ordering only
reflects how evolved an object is along its route from cloud core to
star. Objects of different mass may proceed at different speeds along
this track. No correlation is found between either the point-source flux
or the ratio of the point-source flux over the total 1.1~mm flux, and
the evolutionary phase $\int T_{\rm mb} dV ({\rm HCO^+}\,3\!-\!2) \,/\,
L_{\rm bol}$ of our nine sources. This suggests that the mass of the
disks does not change significantly during the embedded phase. There is
a trend, however, between the point-source flux, the envelope flux, and
the luminosity, indicating that the disk flux primarily depends on the
mass or activity of the central object.

\placefigure{f10}

Myers \& Ladd \markcite{mye93}(1993) proposed a diagram of $L_{\rm
bol}$ versus the bolometric temperature $T_{\rm bol}$ as an extension
to the Hertzsprung--Russell diagram (the so-called BLT~diagram). The
quantity $T_{\rm bol}$ is defined as the temperature of a blackbody
having a maximum at the same frequency as the SED of the source. In
Fig.~10 the BLT~diagram of the TGWW\markcite{tam91} sample is shown
(Chen et~al.\ \markcite{che95}1995), with our subsample again
indicated by filled symbols. It is seen to populate the upper
right-hand region of the diagram, corresponding to young
sources. Although no calculations of evolutionary tracks have been
published for a BLT~diagram yet, a source is expected to proceed from
low $T_{\rm bol}$ and $L_{\rm bol}$ to the main sequence via a maximum
in $L_{\rm bol}$. Compared to our evolutionary tracer, $\int T_{\rm
mb} dV ({\rm HCO^+}\,3\!-\!2) \,/\, L_{\rm bol}$ , which requires the
presence of envelope material, the BLT approach extends over a much
larger range in object age. Our method has the advantage, however,
that mass and evolution can be separated independent of any collapse
model, and that only a sufficiently well-behaved $\dot M_{\rm acc}$ is
required. A more extended survey of HCO$^+$ and H$^{13}$CO$^+$ in a
larger variety of objects will therefore be interesting to further
investigate this scenario.

% -----------------------------------------------------------------

\section{Summary}

The envelopes around a sample of nine embedded YSOs have been
investigated with 3.4 and 2.7~mm continuum interferometry, and with
single-dish observations of the HCO$^+$ and H$^{13}$CO$^+$ 1--0, 3--2
and 4--3 transitions. The sources were selected from the IRAS flux and
color-limited sample defined by TGWW\markcite{tam91}. The results of
this paper can be summarized as follows.

1.~Continuum emission at 3.4 and 2.7~mm is detected in the OVRO beam
toward all nine sources with integrated fluxes ranging between 4~mJy
and 200~mJy. The emission can be described by an unresolved ($<3''$)
point source and, in about half of the objects, emission from an
extended, partially resolved envelope. The point-source emission is
attributed to thermal dust emission from a circumstellar disk, and its
high detection rate suggests that such disks are established already
early in the formation and evolution of protostars. No significant
change in the ratio of disk mass over envelope mass is inferred during
the subsequent duration of the embedded phase.

2.~By extrapolating the point-source fluxes to 1.1~mm, assuming a
spectral index $\alpha=2.5$, the relative contributions of the disk
and the envelope to the single-dish fluxes observed by
Moriarty-Schieven et~al.\ \markcite{mor94}(1994) are estimated. The
point sources typically contribute 30--75\% to the single-dish
flux. The inferred envelope masses within the $19''$ JCMT beam are
0.001--0.26 $M_\odot$, assuming an average dust temperature of 30~K.

3.~HCO$^+$~$J$=1--0 emission maps reveal extended cores ($> 60''$)
superposed on emission from the surrounding cloud. In the 3--2 and
4--3 lines the emission is concentrated toward the source positions
and marginally resolved ($\sim 20''$), as is expected for a power-law
density distribution. The spectra are generally best fit by a narrow
emission line ($\Delta V \approx$~2--3 km\,s$^{-1}$), broad line
wings, and, in a few cases, a narrow self-absorption component. The
H$^{13}$CO$^+$ 3--2 and 4--3 lines are detected toward half of our
sources, with spectra well fit by a single Gaussian at the $V_{\rm
LSR}$ of the HCO$^+$ self-absorption, if present. From observations of
H$^{13}$CO$^+$ 1--0 by Mizuno et~al.\ \markcite{miz94}(1994) and
C$^{18}$O 1--0 by Hayashi et~al.\ \markcite{hay94}(1994), a
beam-averaged HCO$^+$ abundance of $(1.2\pm 0.4) \times 10^{-8}$ with
respect to H$_2$ is derived.

4.~The 1.1~mm envelope flux and the integrated HCO$^+$ line strengths
are well correlated, especially for the higher $J$ transitions. They
can both be modeled with the inside-out collapse model of Shu
\markcite{shu77}(1977) and with simple power-law density distributions
(slopes $p=$~1--3), assuming that the temperature of the dust and the
gas has a power law with slope $-0.4$, and a value of 30~K at a radius
of 1000~AU. Typical densities at $r=1000$~AU of $1\times
10^4$ to $2\times 10^6$~cm$^{-3}$ are inferred. The fact that the models
agree well with the observations indicates that HCO$^+$ 3--2 and 4--3
can be used as envelope tracers, and that no significant depletion of
molecules takes place over $\sim 20''$ scales. However, more detailed
comparison with line-profile calculations are required before
further constraints can be placed on possible collapse models.

5.~Among the remaining objects of the original sample, i.e., those
not selected for our subsample, there is a group that have similar
1.1~mm single-dish fluxes compared to our YSOs in spite of undetected
HCO$^+$ 3--2 emission. It is proposed that all their 1.1~mm flux
originates in a disk rather than in an extended envelope. Available
C$^{18}$O 3--2 and continuum-interferometer data support this
interpretation.

6.~The quantity $\int T_{\rm mb} dV ({\rm HCO^+}\,3\!-\!2) \,/\,
L_{\rm bol}$ is proposed as an evolutionary tracer independent of
intrinsic mass, and measures the ratio of the mass of the central star
to that of the envelope. It is found that those TGWW\markcite{tam91}
sources which are detected in HCO$^+$ 3--2 have a significantly
different distribution in this tracer compared with sources that
remain undetected in HCO$^+$. This supports the interpretation of the
latter group as more evolved objects, and shows that HCO$^+$ is very
well suited as a sensitive probe of the early embedded phase of
low-mass star formation.

% ----------------------------------------------------------------------

\acknowledgments

The authors wish to thank Remo Tilanus, G\"oran Sandell, Fred Baas,
and Floris van~der~Tak for carrying out part of the JCMT
observations. Lee Mundy is acknowledged for useful discussions
concerning the analysis of the continuum observations. The staffs of
OVRO, JCMT and IRAM~30m are thanked for support during various
observing runs. MRH is indebted to the Caltech Divisions of Geological
and Planetary Sciences and of Mathematics, Physics and Astronomy, as
well as the Owens Valley Radio Observatory for their hospitality. The
{\it Stimuleringsfonds Internationalisering\/} of the Netherlands
Organization for Scientific Research (NWO) and the Leids
Kerkhoven--Bosscha Fonds provided travel support for MRH. Research in
Astrochemistry in Leiden is supported by NWO/NFRA through grant
no.~781--76--015.  GAB gratefully acknowledges support provided by
NASA grants NAGW--2297 and NAGW--1955, and HJvL support by the
European Union under contract CHGECT920011. The referee is thanked for
providing insightful comments, which led to a much improved revised
manuscript.

% ----------------------------------------------------------------------

%%% 4. (no APPENDICES)

% ----------------------------------------------------------------------

%%% 5. REFERENCES

% ----------------------------------------------------------------------

%%% 6. FIGURE CAPTIONS

\newpage

\figcaption[fig1.eps]{
HCO$^+$~\hbox{3--2} spectra of the sources defined by Tamura et~al.\
(1991, TGWW) as YSOs (see Table~2). The nine strongest objects
selected for further study are shown in the top panel; the others are
shown in the lower panel. The vertical scale is antenna temperature
$T_{\rm mb}$ in~K, the horizontal scale is velocity $V_{\rm LSR}$
in~\hbox{km\,s$^{-1}$}. The estimated calibration uncertainty of these
HCO$^+$ 3--2 spectra is at least 30\%, as discussed in \S 3.2.
\label{f1}}

\figcaption[fig2a.eps,fig2b.eps]{
In the left-hand panels the vector-averaged, $uv$-binned 3.4~and
2.7~mm visibility amplitudes in~mJy are plotted against $uv$
separation in~k$\lambda$ . The observations, indicated by the filled
symbols, are shown with their $1\sigma$~error bars. The dotted line is
the zero-signal expectation value, the thick solid line a model fit
(see \S 4.1, Table~4). In the right-hand panels the cleaned images of
the 3.4~and 2.7~mm continuum emission are shown, using natural
weighting. The beam~sizes are indicated in the lower left corner of
each panel. The contour levels start at $2\sigma$ ($\approx
3$~mJy\,bm$^{-1}$; except for T~Tau: $\sim 10$~mJy\,bm$^{-1}$ at
3.4~mm, $\sim 6$~mJy\,bm$^{-1}$ at 2.7~mm), and increase in steps of
$2\sigma$.
\label{f2}}

\figcaption[fig3.eps]{
Spectra of HCO$^+$~\hbox{1--0} (bottom),~\hbox{3--2} (middle),
and~\hbox{4--3} (top curves) toward the nine YSOs of our sample (solid
lines). The H$^{13}$CO$^+$~\hbox{3--2} and~4-3 spectra are indicated
by the dotted lines. The \hbox{3--2} and \hbox{4--3} spectra are
offset by 10 and 20~K respectively; the H$^{13}$CO$^+$ spectra have
been multiplied by a factor of~5 (factor of~10 for L1489~IRS and
T~Tau; factor of~2.5 for L1551~IRS~5, as indicated). The vertical
scale is antenna temperature $T_{\rm mb}$ in~K, the horizontal scale
is velocity $V_{\rm LSR}$ in~\hbox{km\,s$^{-1}$}.
\label{f3}}

\figcaption[fig4a.eps,fig4b.eps,fig4c.eps]{
({\it a\/})~Contour maps of integrated HCO$^+$~\hbox{1--0} (left-hand
panels), \hbox{3--2}~(upper right-hand panels), and \hbox{4--3}~(lower
right-hand panels) of L1489~IRS, Haro~6--10, L1551~IRS~5, and
L1535~IRS. The spectra have been integrated over the full velocity
extent of the lines. The positions at which spectra were obtained are
indicated by dots, unless the maps are fully sampled at half
beam-width intervals. The first contour level and the level increment
are $4\sigma$ for 1--0, and $2\sigma$ for 3--2 and 4--3.
(\hbox{1--0}: 1.0~\hbox{K\,km\,s$^{-1}$} for L1489~IRS, Haro~6--10,
and L1535~IRS; 3.0~\hbox{K\,km\,s$^{-1}$} for
L1551~IRS~5. \hbox{3--2}: 1.0~\hbox{K\,km\,s$^{-1}$}. \hbox{4--3}:
1.5~\hbox{K\,km\,s$^{-1}$} for L1489~IRS, Haro~6--10, and L1535~IRS;
4.5~\hbox{K\,km\,s$^{-1}$} for L1551~IRS). The 50\% intensity contour
is indicated by the thick solid line. The respective beam sizes are
shown in the lower left corner of each panel.  ({\it b\/})~Same for
TMR~1, TMC~1A, L1527~IRS, and TMC~1 (\hbox{1--0}:
$4\sigma=1.0$~\hbox{K\,km\,s$^{-1}$} for TMR~1, TMC~1A, and TMC~1;
3.0~\hbox{K\,km\,s$^{-1}$} for L1527~IRS. \hbox{3--2}:
$2\sigma=1.0$~\hbox{K\,km\,s$^{-1}$}. \hbox{4--3}:
$2\sigma=1.5$~\hbox{K\,km\,s$^{-1}$}.)  ({\it c\/})~Same for
T~Tau. Note that a region of only $20''\times 20''$ was mapped in
\hbox{1--0}. (\hbox{1--0}: $4\sigma=3.0$~\hbox{K\,km\,s$^{-1}$}. \hbox{3--2}:
$2\sigma=3.0$~\hbox{K\,km\,s$^{-1}$}. \hbox{4--3}:
$2\sigma=4.5$~\hbox{K\,km\,s$^{-1}$}.)
\label{f4}}

\figcaption[fig5.eps]{
Integrated HCO$^+$ (top panels) and H$^{13}$CO$^+$ (lower panels)
\hbox{1--0} (left), \hbox{3--2} (middle), and \hbox{4--3} (right)
emission as functions of 1.1~mm envelope flux. The objects of our
sample are indicated by the symbols, together with error bars
corresponding to the $\sim 30\%$ calibration uncertainty. The error
bar for T~Tau in the HCO$^+$ 4--3 panel also contains the large
contribution from the outflow to the line wings. Curves are drawn for
the three power-law models, and for the Shu~(1977) collapse model. In
the HCO$^+$~3--2 panel the objects are identified, for comparison with
Fig.~8.
\label{f5}}

\figcaption[fig6.eps]{
({\it a\/})~Continuum flux at 1.1~mm in mJy as function of
post-collapse time in yr for the Shu~(1977) collapse model. The time
at which the collapse expansion wave reaches the outer boundary of the
model core is indicated by the dashed line.  ({\it b\/})~1.1~mm flux
in mJy as function of H$_2$ number density at an arbitrary radius of
1000~AU, for the Shu~(1977) collapse model and three power-law density
distributions with slopes $p=1$,~2 and~3.  ({\it c\/})~Integrated
intensity of HCO$^+$~\hbox{4--3} (top), \hbox{3--2} (middle), and
\hbox{1--0} (bottom), in~\hbox{K\,km\,s$^{-1}$}, as functions of
density at 1000~AU for the same models, assuming a constant HCO$^+$
abundance of $1.2\times 10^{-8}$.  ({\it d\/})~FWHM source size, after
beam convolution, for HCO$^+$~\hbox{4--3} (top), \hbox{3--2} (middle),
and \hbox{1--0} (bottom) as functions of density at 1000~AU for the
same models. The respective beam sizes are indicated by the thick
horizontal lines.
\label{f6}}

\figcaption[fig7.eps]{
Integrated intensity in CS~\hbox{3--2}, 5--4 (CSO and JCMT), and 7--6
(from Moriarty-Schieven et~al.\ 1995) against HCO$^+$~\hbox{3--2} for the
TGWW source sample. The nine objects investigated in the present paper
are indicated by the filled symbols. The size of the symbols
corresponds to the uncertainty in the data.
\label{f7}}

\figcaption[fig8.eps]{
Integrated HCO$^+$~\hbox{3--2} against 1.1~mm single-dish flux for the
full source sample. The nine objects investigated in the present paper
are indicated by the filled symbols. The size of the symbols reflects
the uncertainty in the data. Note that, in contrast with the fluxes
used in Fig.~5, no correction is made for the contribution of any
compact disks. For reference, the curve corresponding to the
Shu~(1977) collapse model is plotted.
\label{f8}}

\figcaption[fig9.eps]{
Cumulative distribution of $\int T_{\rm mb} dV ({\rm HCO^+}\,3\!-\!2)
\,/\, L_{\rm bol}$ of the selected source sample (with and without
correction for line opacity), those sources in the TGWW sample which
are weak in HCO$^+$~\hbox{3--2}, and the sources undetected in
HCO$^+$~\hbox{3--2}. The sources in the selected sample are identified
along the $\tau$-corrected curve. The quantity $\int T_{\rm mb} dV
({\rm HCO^+}\,3\!-\!2) \,/\, L_{\rm bol}$ traces $M_{\rm
env}/M_\star$, i.e., the relative evolutionary phase of an embedded
object. Relatively young sources are located on the right of the plot,
more evolved ones on the left.
\label{f9}}

\figcaption[fig10.eps]{
Diagram of bolometric luminosity against bolometric temperature
(BLT-diagram) of the TGWW source sample. The objects studied in the
present paper are indicated by the filled symbols. The main
sequence is located at the extreme left of the plot.
\label{f10}}

% ----------------------------------------------------------------------

%%% 7. TABLES 1-6

%%% table 1
 
\begin{deluxetable}{lrrrcrrrrrr}
\tablecolumns{11}
\tablewidth{0pt}
\scriptsize
\tablecaption{Selected source sample\label{t1}}
\tablehead{
\colhead{Source} & \colhead{IRAS PSC} & 
\colhead{$\alpha$ (1950.0)\tablenotemark{\,a}} & 
\colhead{$\delta$ (1950.0)\tablenotemark{\,a}} & 
\colhead{Visible/} & \colhead{K} & 
\colhead{NIR\tablenotemark{\,b}} &
\colhead{$F_{100\,\mu{\rm m}}$} & 
\colhead{IRAS\tablenotemark{\,c}} & 
\colhead{$L_{\rm bol}$} &
\colhead{$M_\star$\tablenotemark{\,d}} \nl
 & & 
\colhead{h\ \ m\ \ ss.s} & \colhead{\arcdeg\ \ \arcmin\ \ \arcsec} &
\colhead{Embedded} & \colhead{mag} & \colhead{slope} &
\colhead{Jy} & \colhead{color} & \colhead{$L_{\sun}$} & \colhead{$M_\odot$}
}
\startdata
L1489~IRS  & 04016+2610 & 04\ \ 01\ \ 40.5 & +26\ \ 10\ \ 48 &
   Emb & 9.3 & 2.0 & 56.0 & $-0.49$ & 3.70 & 0.4\phn \nl
T~Tau      & 04190+1924 & 04\ \ 19\ \ 04.1 & +19\ \ 25\ \ 06 &
   Vis\tablenotemark{\,e} & (5.4) & 0.9 & 98.1 & $-0.35$ & 
   25.50\tablenotemark{\,f} & 2.7\phn \tablenotemark{f} \nl
Haro~6--10 & 04263+2426 & 04\ \ 26\ \ 21.9 & +24\ \ 26\ \ 29 &
   Vis\tablenotemark{\,e} & 7.6 & 2.6 & 49.3 & $-0.19$ & 6.98 & 0.9\phn \nl
L1551~IRS~5 & 04287+1801 & 04\ \ 28\ \ 40.2 & +18\ \ 01\ \ 42 &
   Emb & 9.3 & 2.8 & 457.9 & $-0.55$ & 21.90 & 2.6\phn \nl
L1535~IRS  & 04325+2402 & 04\ \ 32\ \ 33.4 & +24\ \ 02\ \ 13 &
   Emb & 11.1 & 1.8 & 23.0 & $-0.79$ & 0.70 & 0.15 \nl
TMR~1     & 04361+2547 & 04\ \ 36\ \ 09.7 & +25\ \ 47\ \ 29 &
   Emb & 10.6 & 2.5 & 33.1 & $-0.38$ & 2.90 & 0.3\phn \nl
TMC~1A    & 04365+2535 & 04\ \ 36\ \ 31.1 & +25\ \ 35\ \ 54 &
   Emb & 10.6 & 2.2 & 38.0 & $-0.62$ & 2.20 & 0.3\phn \nl
L1527~IRS   & 04368+2557 & 04\ \ 36\ \ 49.6 & +25\ \ 57\ \ 21 &
   Emb & (13.0) & 2.1 & 71.0 & $-1.38$ & 1.30 & 0.2\phn \nl
TMC~1     & 04381+2540 & 04\ \ 38\ \ 08.4 & +25\ \ 40\ \ 52 &
   Emb & 12.0 & 2.3 & 12.6 & $-0.58$ & 0.66 & 0.15 \nl
\enddata
\tablenotetext{a}{\,Best fit positions to 3.4 and 2.7~mm continuum
   interferometric data (see \S 4.1).}
\tablenotetext{b}{\,Near-infrared spectral slope, defined as $s = d\log
   S_\nu\,/\,d\log\lambda$ between 2.2 and 25~$\mu$m.}
\tablenotetext{c}{\,IRAS color, defined as
   $\log\,\bigl[F_\nu(25\,\mu{\rm m}) \,/\,F_\nu(60\,\mu{\rm
   m})\bigr]$.}  
\tablenotetext{d}{\,Maximum mass of central object, assuming that all
   bolometric luminosity is stellar and that the object is on the
   birthline.}  
\tablenotetext{e}{\,With embedded companion.}  
\tablenotetext{f}{\,Sum of T~Tau~N and~S.}
\tablerefs{
Cohen, Emerson \& Beichman~\markcite{coh89}1989  ($L_{\rm bol}$ T~Tau); 
Leinert \& Haas~\markcite{lei89}1989 (K~photometry Haro~6--10); 
Tamura et~al.~\markcite{tam91}1991 (K~photometry, NIR~slope);
Kenyon \& Hartmann~\markcite{key95}1995 ($L_{\rm bol}$).
}
\end{deluxetable}

%%% table 2.

\begin{deluxetable}{llcrrrrrr}
\scriptsize
\tablecolumns{9}
\tablewidth{0pt}
\tablecaption{Full source sample\label{t2}}
\tablehead{
\colhead{IRAS PSC} & \colhead{Name} & \colhead{Visual/} &
  \colhead{$T_{\rm eff}$} & \colhead{$L_\star$} & \colhead{$T_{\rm bol}$} &
  \colhead{$L_{\rm bol}$} & \colhead{$F_\nu\,(1.1\,{\rm mm})$\tablenotemark{~a}} &
  \colhead{HCO$^+$ 3--2\tablenotemark{~a}} \cr
 & & \colhead{Embedded} & \colhead{K} & \colhead{$L_\odot$} &
  \colhead{K} & \colhead{$L_\odot$} & \colhead{Jy} & 
  \colhead{K\,km\,s$^{-1}$}}
\startdata
04016+2610 & L1489~IRS & Emb & \nodata & \nodata & 238 & 3.70 & $0.180 \pm 0.021$ & $6.90 \pm 0.40$\nl
04108+2803 & & Emb & \nodata & \nodata & 205 & 0.72 & $<0.1$ & $<0.25$ \nl
04113+2758 & & Emb & \nodata & \nodata & 606 & 2.0 & $0.461 \pm 0.053$ & $<0.29$ \nl
04169+2702 & & Emb & \nodata & \nodata & 170 & 0.80 & $0.281 \pm 0.053$ & $2.10 \pm 0.09$ \nl
04181+2655 & & Emb & \nodata & \nodata & 278 & 0.43 & $0.044 \pm 0.026$ & $1.10 \pm 0.15$ \nl
04190+1924 & T~Tau & Vis+Emb & \nodata & \nodata & \nodata & 25.50\tablenotemark{~b} & $0.579 \pm 0.027$ & $17.90 \pm 0.50$ \nl
04191+1523 & & Emb & \nodata & \nodata & \nodata & 0.48 & $0.179 \pm 0.027$ & $0.97 \pm 0.09$ \nl
04239+2436 & & Emb & \nodata & \nodata & 236 & 1.27 & $0.114 \pm 0.021$ & $0.82 \pm 0.06$ \nl
04240+2559 & DG~Tau & Vis & \nodata & \nodata & 1440 & 6.36 & $0.523 \pm 0.048$ & $<0.20$ \nl
04248+2612 & HH~31~IRS & Emb & \nodata & \nodata & 334 & 0.36 & $0.099 \pm 0.015$ & $1.90 \pm 0.08$ \nl
04263+2426 & Haro~6--10 & Vis+Emb & 4730 & \nodata & 253 & 6.98 & $0.111 \pm 0.011$ & $2.42 \pm 0.11$ \nl
04287+1801 & L1551~IRS~5 & Emb & \nodata & \nodata & 97 & 21.90 & $2.77\phn \pm 0.30\phn$ & $9.64 \pm 0.22$ \nl
04288+2417 & HK~Tau & Vis & 3785 & 1.30 & 2148 & 0.81 & $0.110 \pm 0.020$ & $<0.18$ \nl
04292+2422 & Haro~6--13 & Vis & \nodata & \nodata & 910 & 1.30 & $0.233 \pm 0.022$ & $<0.19$ \nl
04295+2251 & L1536~IRS & Emb & \nodata & \nodata & 447 & 0.44 & $0.094 \pm 0.018$ & $<0.19$ \nl
04296+1725 & GG~Tau & Vis & 4060 & 1.31 & 2621 & 2.00 & $0.74\phn \pm 0.12\phn$ & $<0.26$ \nl
04302+2247 & & Emb & \nodata & \nodata & 202 & 0.34 & $0.149 \pm 0.019$ & $<0.19$ \nl
04325+2402 & L1535~IRS & Emb & \nodata & \nodata & 157 & 0.70 & $0.074 \pm 0.015$ & $2.30 \pm 0.13$ \nl
04328+2248 & HP~Tau & Vis & 4730 & 1.22 & 2748 & 2.40 & $<0.1$ & $<0.18$ \nl
04361+2547 & TMR~1 & Emb & \nodata & \nodata & 144 & 2.90 & $0.188 \pm 0.027$ & $8.20 \pm 0.40$ \nl
04365+2535 & TMC~1A & Emb & \nodata & \nodata & 172 & 2.20 & $0.438 \pm 0.038$ & $3.30 \pm 0.40$ \nl
04368+2557 & L1527~IRS & Emb & \nodata & \nodata & 59 & 1.30 & $0.482 \pm 0.037$ & $8.80 \pm 0.20$ \nl
04381+2540 & TMC~1 & Emb & \nodata & \nodata & 139 & 0.66 & $0.116 \pm 0.013$ & $3.20 \pm 0.30$ \nl
04390+2517 & LkH$\alpha$~332/G2 & Vis & 4060 & 1.94 & 2563 & 1.60 & $<0.1$ & $<0.20$ \nl
\enddata
\tablenotetext{a}{\,In $19''$ beam.}
\tablenotetext{b}{\,Sum of T~Tau~N and~S. N has $L_{\rm
bol}=15.50$~$L_\odot$, $L_\star = 7.09$~$L_\odot$, $T_{\rm bol} = 3452$~K,
and $T_{\rm eff} = 5250$~K; S has $L_{\rm bol} = 10.0$~$L_\odot$ and
$T_{\rm bol} = 501$~K.}
\tablerefs{
Cohen et al.~\markcite{coh89}1989 ($L_\star$); 
Tamura et al.~\markcite{tam91}1991 (source sample); 
Moriarty-Schieven et~al.~\markcite{mor94}1994 ($F_\nu$); 
Kenyon \& Hartmann~\markcite{key95}1995 ($T_{\rm eff}$, $L_{\rm bol}$); 
Chen et~al.~\markcite{che95}1995,~\markcite{che97}1997 ($T_{\rm bol}$, 
   $L_{\rm bol}$ 04113+2658 and LkH$\alpha$ 332/G2); 
Ohashi et~al.~\markcite{oha96}1996 ($L_{\rm bol}$ 04191+1524).
}
\end{deluxetable}

%%% table 3
 
\begin{deluxetable}{llll}
\scriptsize
\tablecolumns{4}
\tablewidth{0pt}
\tablecaption{Overview of observations\label{t3}}
\tablehead{
\colhead{Year/Month} & \colhead{Instrument} & \colhead{Observation} &
\colhead{Sources}}
\startdata
1992/4, 1993/7 &
        OVRO    & 
        $F_\nu(\lambda=3.4\,{\rm mm})$  &
        T~Tau \nl
1993/10, 1994/2--4 &
        OVRO    & 
        $F_\nu(\lambda=3.4\,{\rm mm})$  &
        all\,\tablenotemark{a}\ , except T~Tau \nl
1993/1,6 &
        OVRO &
        $F_\nu(\lambda=2.7\,{\rm mm})$  &
        T~Tau \nl
1995/2--5 &
        OVRO &
        $F_\nu(\lambda=2.7\,{\rm mm})$  &
        Haro~6--10, L1551~IRS~5, L1535~IRS, TMR~1, L1527~IRS, TMC~1 \nl
1996/10, 1997/2 &
        OVRO &
        $F_\nu(\lambda=2.7\,{\rm mm})$  &
        L1489~IRS, TMC~1A \nl
1991/5 &
        IRAM 30m &
        HCO$^+$ 1--0 &
        T~Tau \nl
1995/5 &
        IRAM 30m &
        HCO$^+$ 1--0 &
        all\,\tablenotemark{a}\ , except T~Tau \nl
1993/8 &
        JCMT &
        HCO$^+$ 3--2 &
        full source sample (see Table~2) \nl
1995/8 &
        JCMT &
        HCO$^+$ 3--2 &
        L1489~IRS, T~Tau, TMR~1, TMC~1A, TMC~1 \nl
1994/12 &
        JCMT &
        HCO$^+$ 4--3 &
        all\,\tablenotemark{a}\nl
1994/12, '95/10, '96/5 &
        JCMT &
        H$^{13}$CO$^+$ 3--2, 4--3 &
        all\,\tablenotemark{a}\nl
\enddata
\tablenotetext{a}{\,L1489~IRS, T~Tau, Haro~6--10, L1551~IRS~5, L1535~IRS, TMR~1, TMC~1A, L1527~IRS and TMC~1.}
\end{deluxetable}

%%% table 4
 
\begin{deluxetable}{lrrrrrrrr}
\scriptsize
\tablecolumns{8}
\tablewidth{0pt}
\tablecaption{Fit parameters to millimeter continuum visibilities\label{t4}}
\tablehead{
 & \multicolumn{3}{c}{$\lambda=3.4$ mm} & &
\multicolumn{3}{c}{$\lambda=2.7$ mm} \nl
\cline{2-4} \cline{6-8} \nl
\colhead{Source} & \colhead{$F_\nu^{\rm \,point}$} & 
\colhead{$F_\nu^{\rm \,Gaussian}$} & \colhead{FWHM} & 
& \colhead{$F_\nu^{\rm \,point}$} &  \colhead{$F_\nu^{\rm \,Gaussian}$} & 
\colhead{FWHM} & \colhead{$M_{\rm disk}$\tablenotemark{\,a}} \nl
\colhead{} & \colhead{mJy} & \colhead{mJy} & \colhead{\arcsec} &
& \colhead{mJy} & \colhead{mJy} & \colhead{\arcsec} & \colhead{$M_\odot$} 
}
\startdata
L1489~IRS       & $4.3 \pm 1.1$\tablenotemark{\,b} & \nodata  & \nodata & 
                & $6.8 \pm 1.3$\tablenotemark{\,b} & \nodata  & \nodata & $(4.4\pm 0.7)\times 10^{-3}$ \cr
T~Tau           & $45.6 \pm 5.5$ & \nodata & \nodata & 
                & $39.1 \pm 4.2$ & $35.3 \pm 10.4$ & $9\times 7$ & $(2.3 \pm 0.3)\times 10^{-2}$ \cr
Haro~6--10      & $11.9 \pm 1.7$& \nodata & \nodata &
                & $11.3 \pm 1.4$ & $13.7 \pm\phn 4.3$ & $7\times 7$ & $(1.0 \pm 0.2)\times 10^{-2}$ \cr
L1551~IRS~5     & $80.8 \pm 3.0$ & $48.3 \pm 10.4$ & $10\times 9$ &
                & $97.1 \pm  2.6$ & $99.8 \pm\phn 9.3$ & $9\times 7$ & $(7.3 \pm 0.2) \times 10^{-2}$ \cr
L1535~IRS       & $4.6 \pm 1.1$\tablenotemark{\,b} & \nodata  & \nodata &
                & $6.9 \pm 1.3$\tablenotemark{\,b} & \nodata  & \nodata & $(4.6 \pm 0.7) \times 10^{-3}$ \cr
TMR~1           & $10.0 \pm 1.0$ & \nodata & \nodata &
                & $12.0 \pm 1.2$ & $10.8 \pm\phn 7.6$ & $21\times 3$ & $(9.1 \pm 0.6)\times 10^{-3}$ \cr
TMC~1A          & $18.1 \pm 1.2$ & $(21.2 \pm 21.1)$ & $(56\times 15)$ &
                & $33.1 \pm 1.3$ & \nodata & \nodata & $(2.0 \pm 0.6)\times 10^{-2}$ \cr
L1527~IRS       & $16.2 \pm 2.2$ & $17.9 \pm 3.3$ & $8\times 1$ &
                & $26.1 \pm 2.8$ & $14.4 \pm\phn 4.9$ & $5\times 1$ & $(1.7 \pm 0.1)\times 10^{-2}$ \cr
TMC~1           & $5.0 \pm 1.1$\tablenotemark{\,b} & \nodata  & \nodata &
                & $7.6 \pm 1.3$\tablenotemark{\,b} & \nodata  & \nodata & $(5.0\pm 0.7)\times 10^{-3}$ \cr
\enddata
\tablenotetext{a}{\,Minimum mass of disk assuming optically thin emission and
a dust temperature of 30~K (see \S 4.3).}
\tablenotetext{b}{\,Doubtful fit: flux close to zero-signal expectation value.}
\end{deluxetable}

%%% table 5.
 
\begin{deluxetable}{lrrrr}
%\scriptsize
\tablecolumns{5}
\tablewidth{0pt}
\tablecaption{Envelope fluxes, masses and densities\label{t5}}
\tablehead{
\colhead{Source} & \colhead{$F_\nu$\tablenotemark{\,a}} & 
   \colhead{$F_\nu^{\rm \, env}$\tablenotemark{\;b}} & 
   \colhead{$M_{\rm env}$\tablenotemark{\,c}} & 
   \colhead{$n_{\rm H_2}$\tablenotemark{\,d}} \cr
 & \colhead{mJy} & \colhead{mJy} & \colhead{$M_{\sun}$} &
   \colhead{cm$^{-3}$}
}
\startdata
L1489~IRS  & $180 \pm\phn 21$   & 112--180\tablenotemark{\,e} & 
        0.016--0.025 & $(0.9$--$1.4)\times 10^5$ \nl 
T~Tau      & $579 \pm\phn 27$   & $211\pm\phn 47 $   & 
        $0.029\pm 0.007$ & $(1.6\pm 0.4)\times 10^5$ \nl
Haro~6--10 & $111 \pm\phn 11$ & $<30$\tablenotemark & 
        $<0.0042$ & $<2\times 10^4$ \nl
L1551~IRS~5 & $2770 \pm 300$ & $1875\pm 300$ & 
        $0.26\phn \pm 0.04\phn$ & $(1.8\pm 0.4)\times 10^6$ \nl
L1535~IRS   & $74 \pm\phn 15$ & 0--74\tablenotemark{\,e} & 
        $<0.01$\phn & $<5.8\times 10^4$ \nl
TMR~1      & $188 \pm\phn 27$ & $47\pm\phn 29 $ & 
        $0.007\pm 0.004$ & $(3.7\pm 2.2)\times 10^4$ \nl
TMC~1A    & $438 \pm\phn 38$   & $130\pm\phn 40$   & 
        $0.018\pm 0.006$ & $(1.0\pm 0.3) \times 10^5$ \nl
L1527~IRS      & $482 \pm\phn 37$   & $223\pm\phn 44$   & 
        $0.031\pm 0.006$ & $(1.7\pm 0.3)\times 10^5$ \nl
TMC~1     & $116 \pm\phn 13$   & 38--116\tablenotemark{\,e}& 
        0.005--0.016 & $(3.0$--$9.0)\times 10^4$ \nl
\enddata
\tablenotetext{a}{\,1.1 mm flux from Moriarty-Schieven 
   et~al.~\markcite{mor94}1994, Kenyon et~al.~\markcite{key93a}1993a.}
\tablenotetext{b}{\,Envelope flux, corrected for contribution 
   from the disk, see Table~4.}
\tablenotetext{c}{\,Mass in $19''$ beam, see \S 4.3.}
\tablenotetext{d}{\,Density at arbitrary radius $r=1000$ AU, see \S6.}
\tablenotetext{e}{\,No constraints on spatial distribution of flux in OVRO beam;
   indicated range corresponds to the assumption that all, respectively none,
    of the OVRO flux originates in a disk.}
\end{deluxetable}

%%% table 6
 
\begin{deluxetable}{lrrrrrrrrr}
\scriptsize
\tablecolumns{10}
\tablewidth{0pt}
\tablecaption{HCO$^+$ line intensities, opacities and source sizes
\label{t6}}
\tablehead{
\colhead{Source} & 
   \colhead{$I_{1-0}$\tablenotemark{\,a}} & 
   \colhead{$\theta_{1-0}$\tablenotemark{\,b}} &
   \colhead{$\bar\tau_{1-0}$\tablenotemark{\,c}} &
   \colhead{$I_{3-2}$} & \colhead{$\theta_{3-2}$} & \colhead{$\bar\tau_{3-2}$} &
   \colhead{$I_{4-3}$} & \colhead{$\theta_{4-3}$} & \colhead{$\bar\tau_{4-3}$}\nl
 & \colhead{K\,km\,s$^{-1}$} & \colhead{$''$} &
 & \colhead{K\,km\,s$^{-1}$} & \colhead{$''$} &
 & \colhead{K\,km\,s$^{-1}$} & \colhead{$''$} 
}
\startdata
\cutinhead{HCO$^+$}
L1489~IRS       & $6.7 \pm 0.2$ & 60 & \nodata
                & $6.9 \pm 0.4$ & 24 & $6.8 \pm 1.2$
                & $10.0 \pm 0.3$ & 17 & $4.0 \pm 0.5$ \nl
T~Tau           & $16.6 \pm 0.2$ & \nodata\tablenotemark{\,h} & \nodata
                & $17.9 \pm 0.5$\tablenotemark{\,d} & 25 & $2.3 \pm 0.4$
                & $37.1 \pm 0.5$\tablenotemark{\,e} & 20 & $3.3 \pm 3.1$ \nl
Haro~6--10      & $2.8 \pm 0.1$ & 90 & \nodata
                & $2.4 \pm 0.1$\tablenotemark{\,f} & \nodata\tablenotemark{\,f} & $<5.3$
                & $5.4 \pm 0.5$ & 23 & $<3.0$ \nl
L1551~IRS~5     & $9.2 \pm 0.3$ & 140 & $12.2 \pm 0.9$
                & $9.6 \pm 0.2$\tablenotemark{\,f} & \nodata\tablenotemark{\,f} & $12.7 \pm 2.2$
                & $22.7 \pm 0.4$ & 22 & $6.8\pm 0.5$ \nl
L1535~IRS       & $7.0 \pm 0.1$ & 110 & $9.0 \pm 0.6$
                & $2.3 \pm 0.1$\tablenotemark{\,f} & \nodata\tablenotemark{\,f} & $<6.3$
                & $3.0 \pm 0.7$ & 24 & $<7.8$ \nl
TMR~1           & $4.3 \pm 0.1$ & 48 & $12.0 \pm 1.2$
                & $8.2 \pm 0.4$ & 30 & $4.1 \pm 0.6$ 
                & $7.3 \pm 0.7$ & 24 & $<1.6$ \nl
TMC~1A          & $2.2 \pm 0.1$ & 75\tablenotemark{\,h} & $20.9 \pm 3.5$
                & $3.3 \pm 0.4$ & 19 & $<6.3$
                & $3.5 \pm 0.4$ & 20\tablenotemark{\,h} & $<2.6$ \nl
L1527~IRS       & $6.5 \pm 0.3$ & 63 & $17.9 \pm 1.6$
                & $8.9 \pm 0.2$\tablenotemark{\,f} & \nodata\tablenotemark{\,f} & $6.7 \pm 0.7$
                & $12.9 \pm 0.7$ & \nodata\tablenotemark{\,i} & $1.3 \pm 0.4$ \nl
TMC~1           & $2.2 \pm 0.1$ & 66 & \nodata 
                & $3.2 \pm 0.3$ & 26 & $<6.3$
                & $5.5 \pm 0.7$ & 19 & $<1.1$ \nl
\cutinhead{H$^{13}$CO$^+$}
L1489~IRS       &\nodata
                &\nodata&\nodata& $0.67 \pm 0.07$
                &\nodata&\nodata& $0.61 \pm 0.05$ &\nodata&\nodata\nl
T~Tau           &\nodata
                &\nodata&\nodata& $0.69 \pm 0.07$
                &\nodata&\nodata& $0.58 \pm 0.05$ &\nodata&\nodata\nl
Haro~6--10      &\nodata
                &\nodata&\nodata& $<0.27$
                &\nodata&\nodata& $<0.23$ &\nodata&\nodata\nl
L1551~IRS~5     &$1.57 \pm 0.05$
                &\nodata&\nodata& $2.41 \pm 0.3\phn$
                &\nodata&\nodata& $2.23 \pm 0.1\phn$ &\nodata&\nodata\nl
L1535~IRS       &$0.89 \pm 0.04$\tablenotemark{\,g} 
                &\nodata&\nodata& $<0.27$
                &\nodata&\nodata& $<0.26$ &\nodata&\nodata\nl
TMR~1           &$0.72 \pm 0.05$ 
                &\nodata&\nodata& $0.50 \pm 0.05$
                &\nodata&\nodata& $<0.20$ &\nodata&\nodata\nl
TMC~1A          &$0.60 \pm 0.04$ 
                &\nodata&\nodata& $<0.27$
                &\nodata&\nodata& $<0.13$ &\nodata&\nodata\nl
L1527~IRS       &$1.56 \pm 0.05$ 
                &\nodata&\nodata& $1.24 \pm 0.2\phn$
                &\nodata&\nodata& $0.35 \pm 0.03$ &\nodata&\nodata\nl
TMC~1           &\nodata 
                &\nodata&\nodata& $<0.27$
                &\nodata&\nodata& $<0.12$ &\nodata&\nodata\nl
\enddata
\tablenotetext{a}{\,$I_{u-l}$: integrated intensity $\int T_{\rm mb}dV$ of transition $u$--$l$.}
\tablenotetext{b}{\,$\theta_{u-l}$: FWHM source size.}
\tablenotetext{c}{\,$\bar\tau_{u-l}$: opacity averaged over line profile assuming an abundance ratio of 65:1 for [HCO$^+$]\,:\,[H$^{13}$CO$^+$].}
\tablenotetext{d}{\,At $(0\arcsec,-5\arcsec)$; no significant contribution from outflow.}
\tablenotetext{e}{\,Contribution of outflow to intensity integrated over 0--14
km\,s$^{-1}$ is $\sim 30$ K\,km\,s$^{-1}$.}
\tablenotetext{f}{\,JCMT August 1993: calibration uncertain by 30\%, no map obtained.}
\tablenotetext{g}{\,No line width given; $\Delta V=0.6$ km\,s$^{-1}$ adopted.}
\tablenotetext{h}{\,Extent of emission ill defined.}
\tablenotetext{i}{\,No map obtained.}
\tablerefs{
Mizuno et~al.~\markcite{miz94}1994 (H$^{13}$CO$^+$ 1--\,0, 
   assuming $\eta_{\rm mb}=1.0$).
}
\end{deluxetable}

% ----------------------------------------------------------------------

\newpage
\input psfig

\null
\rightline{\tt\underbar{Fig.~1}}
\centerline{\psfig{figure=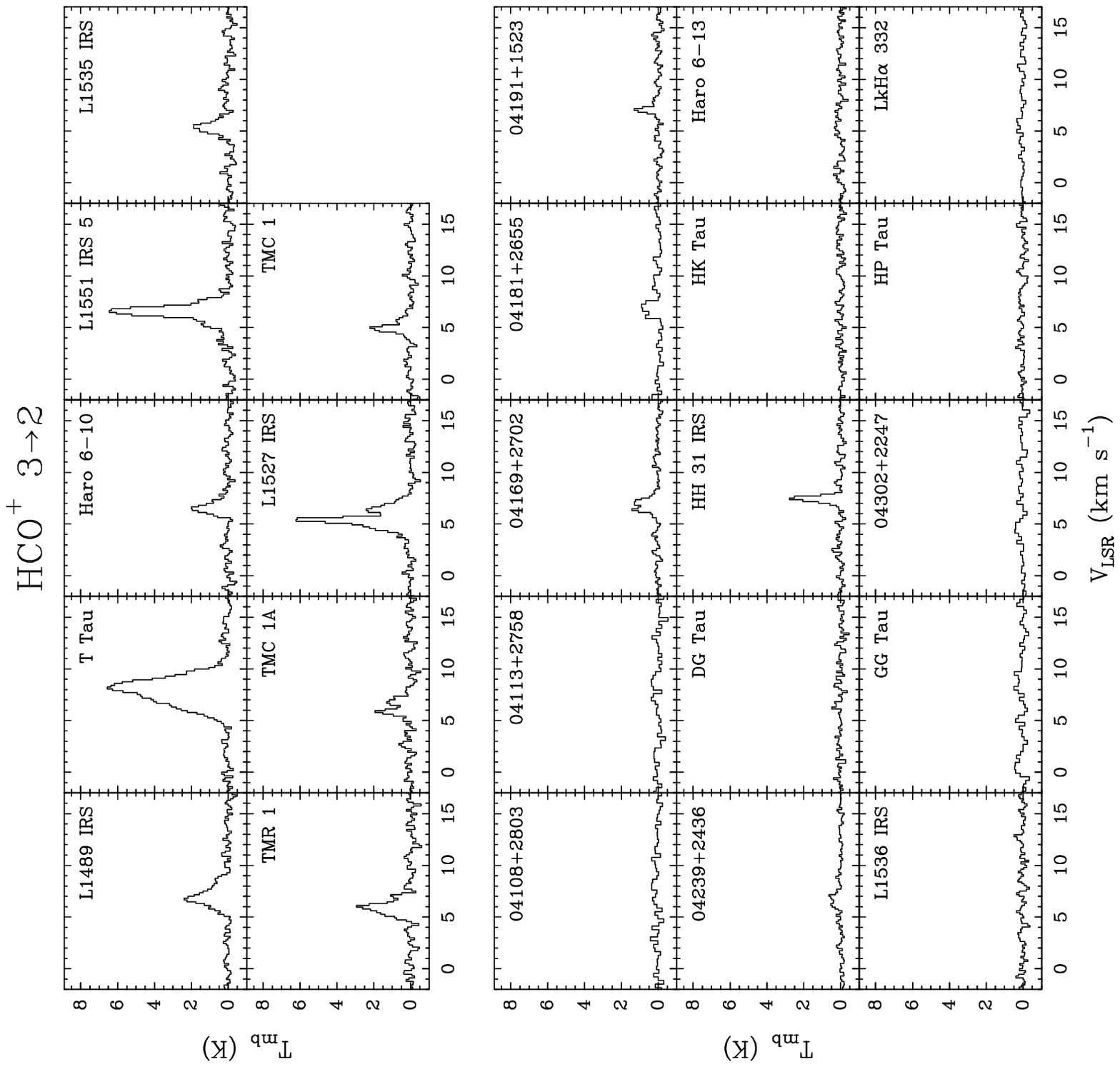,rheight=10truecm,height=25truecm}}
\vfill\eject

\null
\rightline{\tt\underbar{Fig.~2a}}
\centerline{\psfig{figure=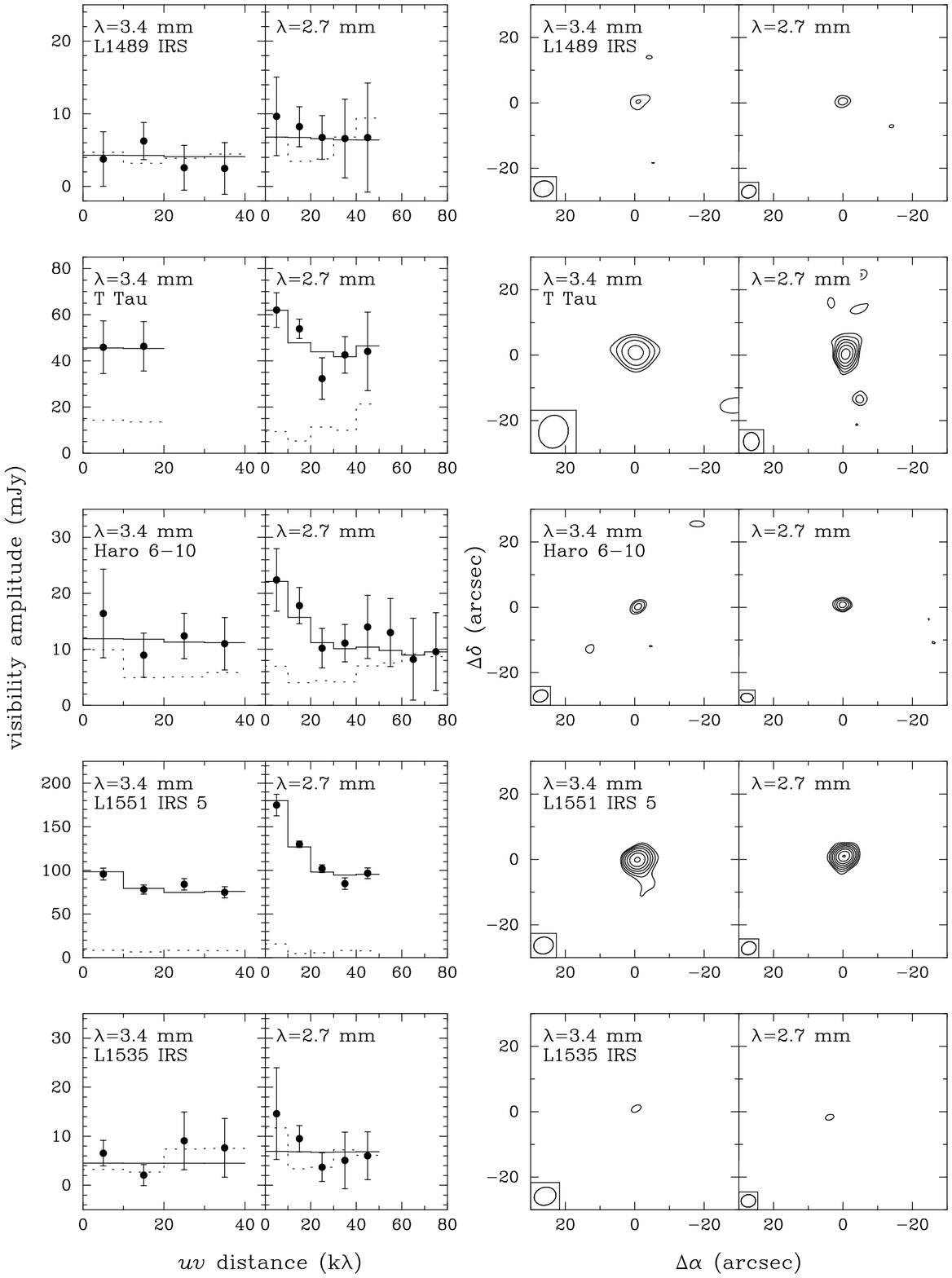,rheight=10truecm,height=25truecm}}
\vfill\eject

\null
\rightline{\tt\underbar{Fig.~2b}}
\centerline{\psfig{figure=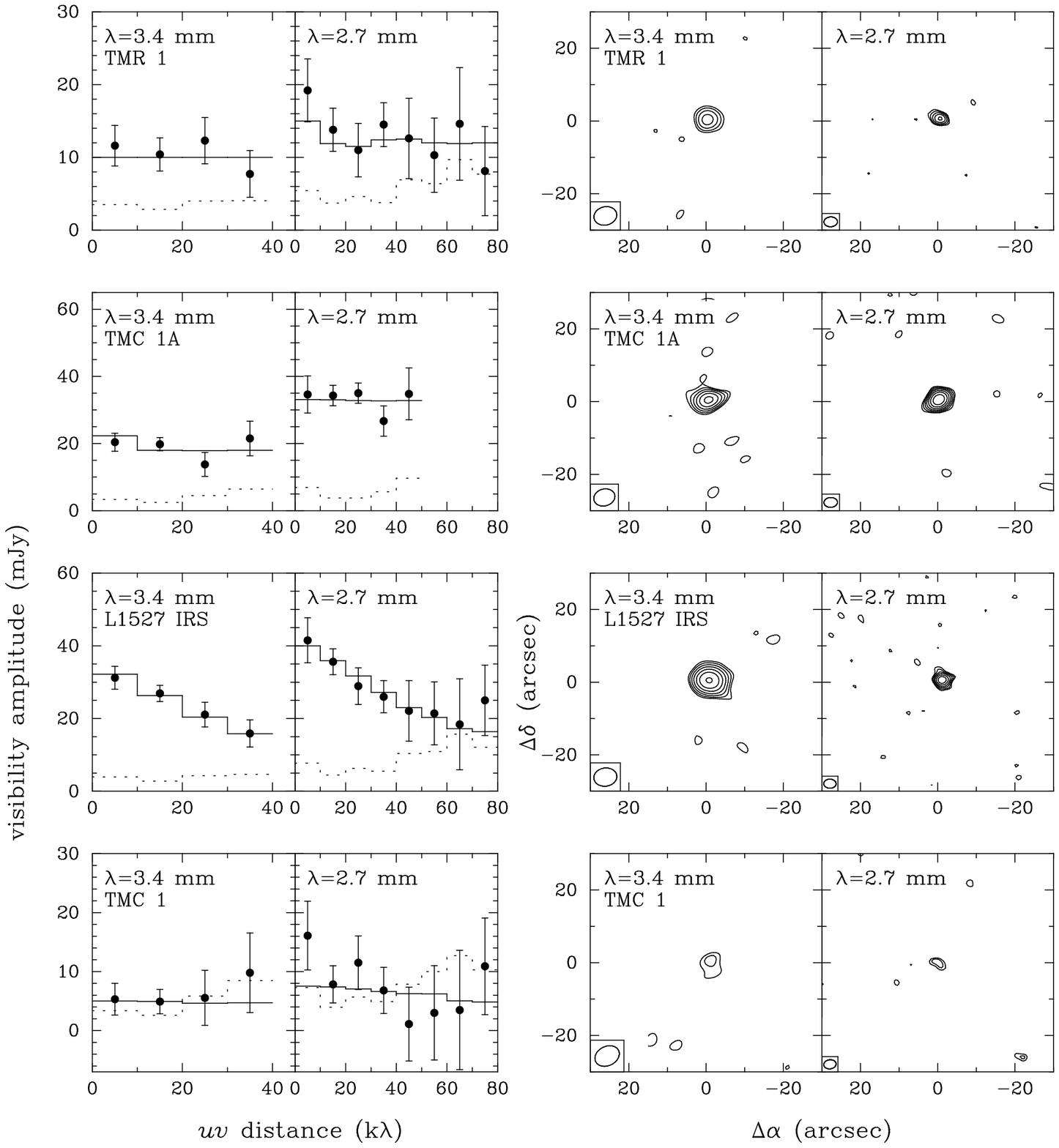,rheight=10truecm,height=25truecm}}
\vfill\eject

\null
\rightline{\tt\underbar{Fig.~3}}
\centerline{\psfig{figure=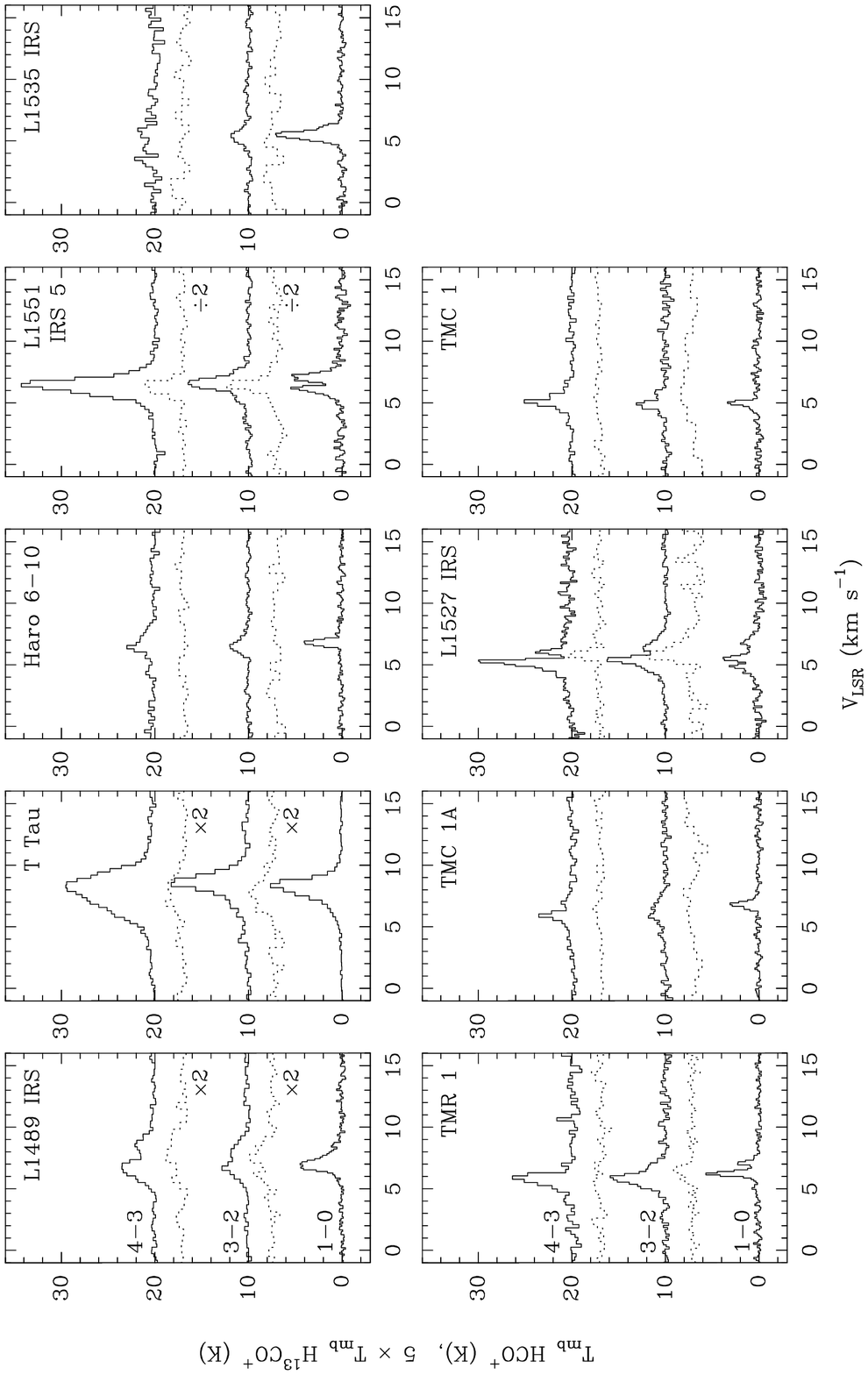,rheight=10truecm,height=25truecm}}
\vfill\eject

\null
\rightline{\tt\underbar{Fig.~4a}}
\centerline{\psfig{figure=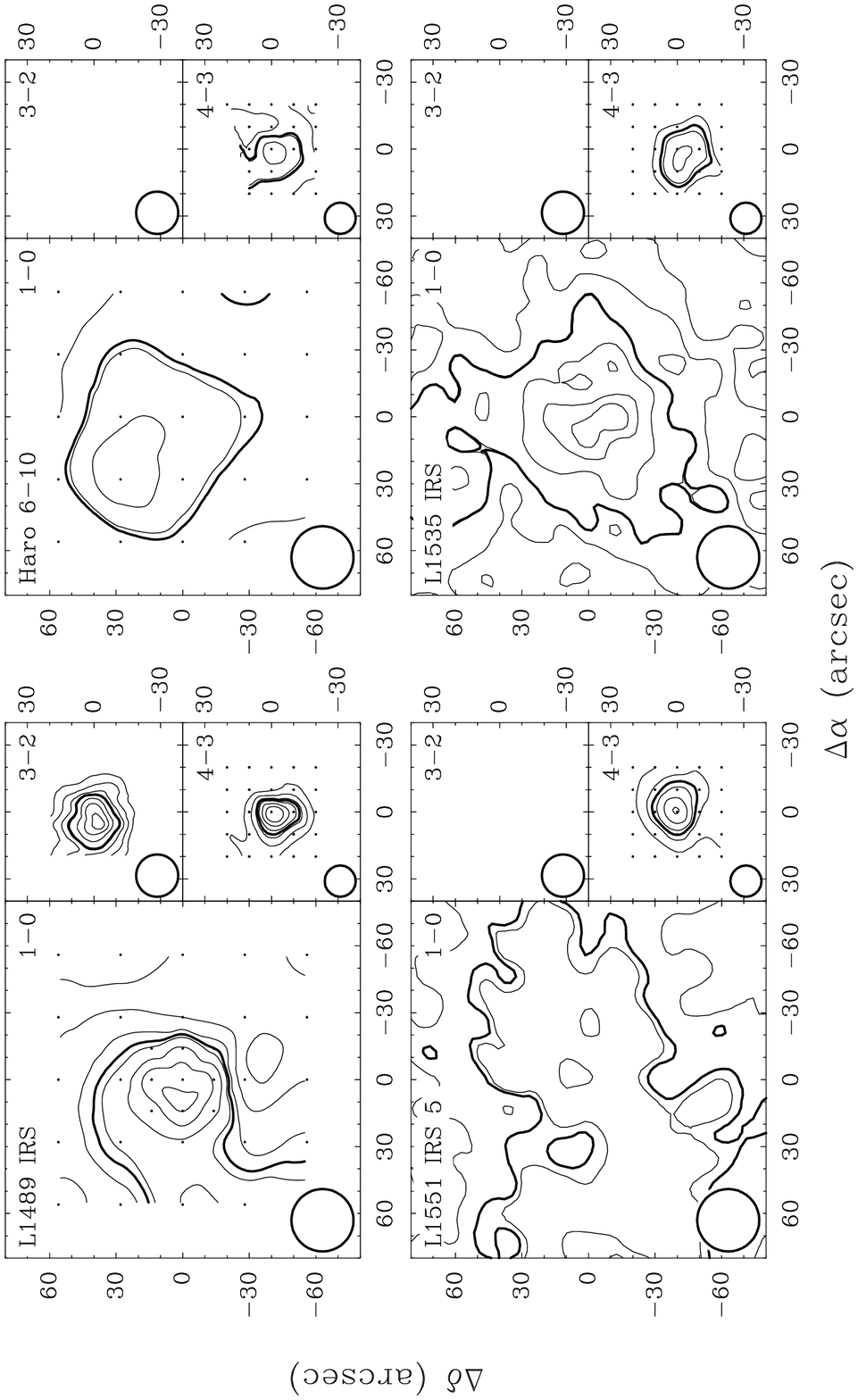,rheight=10truecm,height=25truecm}}
\vfill\eject

\null
\rightline{\tt\underbar{Fig.~4b}}
\centerline{\psfig{figure=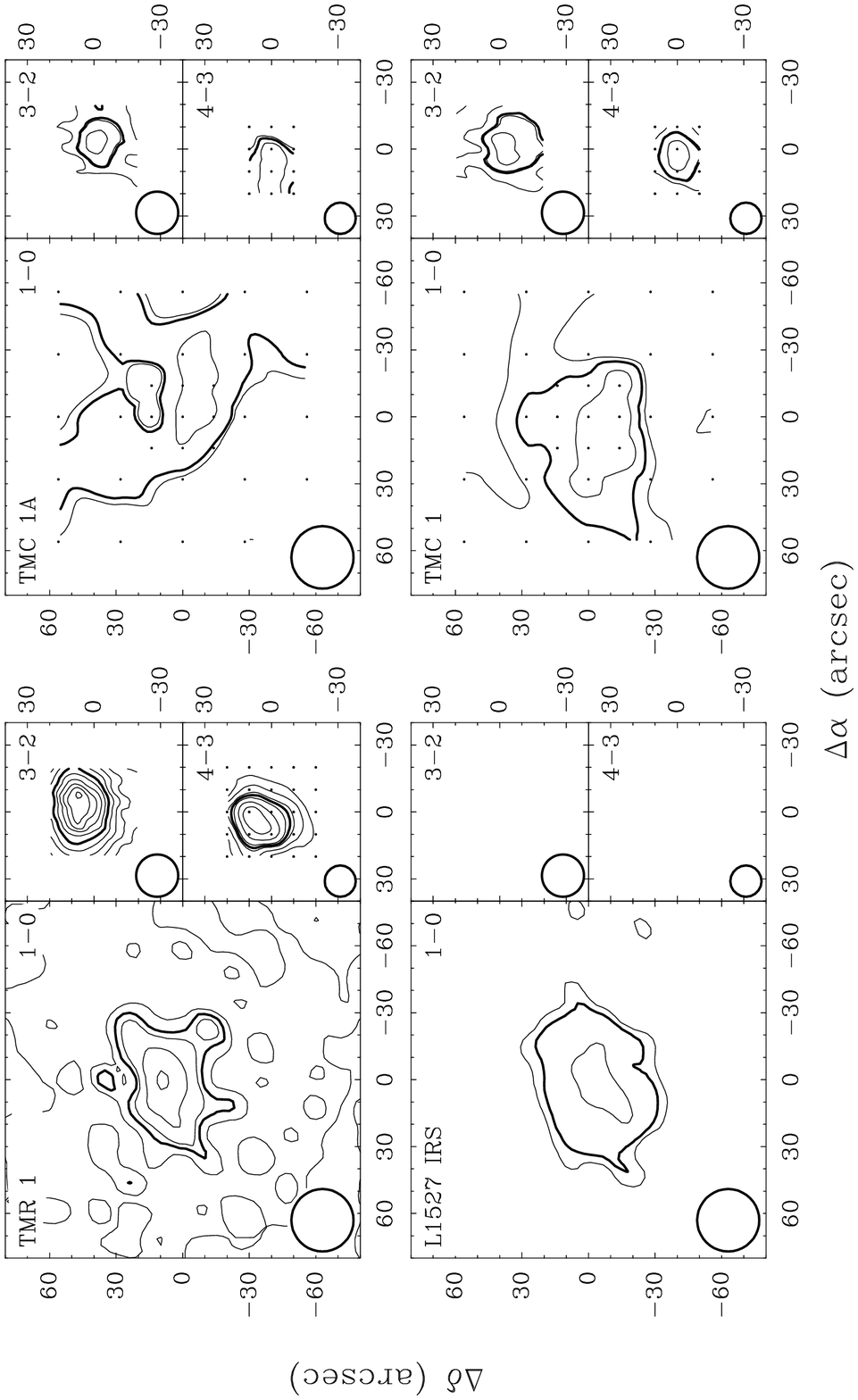,rheight=10truecm,height=25truecm}}
\vfill\eject

\null
\rightline{\tt\underbar{Fig.~4c}}
\centerline{\psfig{figure=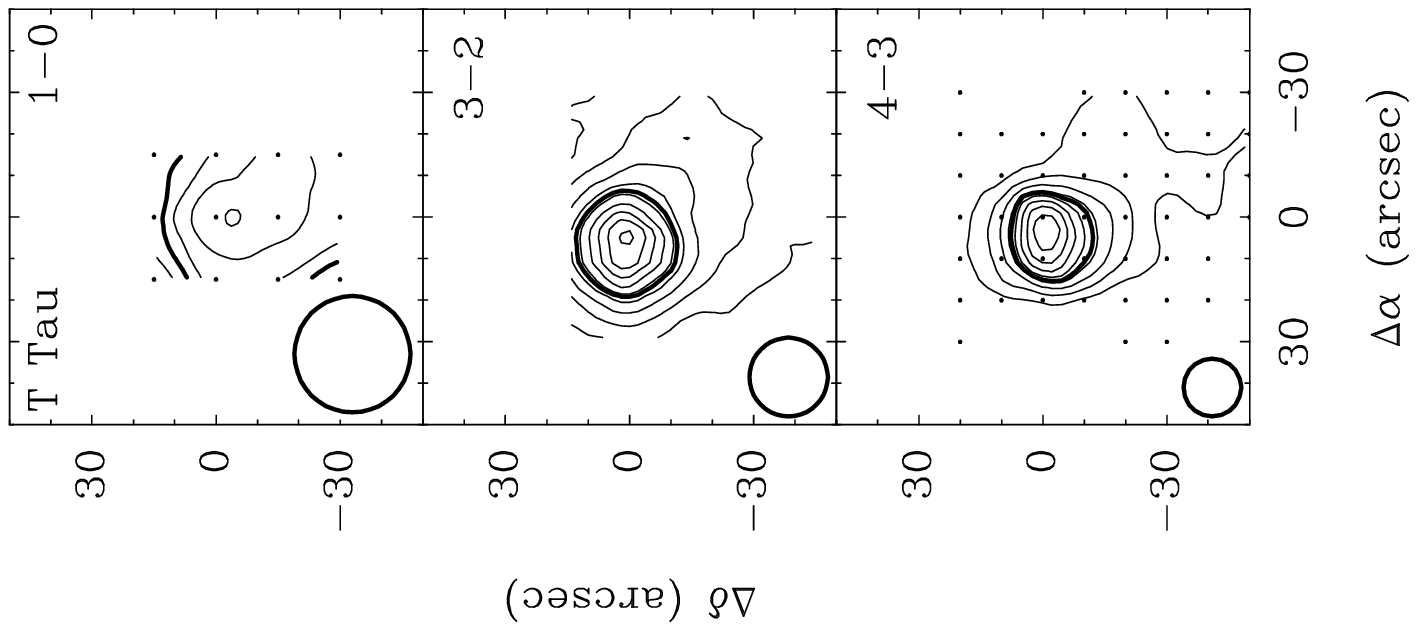,rheight=10truecm,height=25truecm}}
\vfill\eject

\null
\rightline{\tt\underbar{Fig.~5}}
\centerline{\psfig{figure=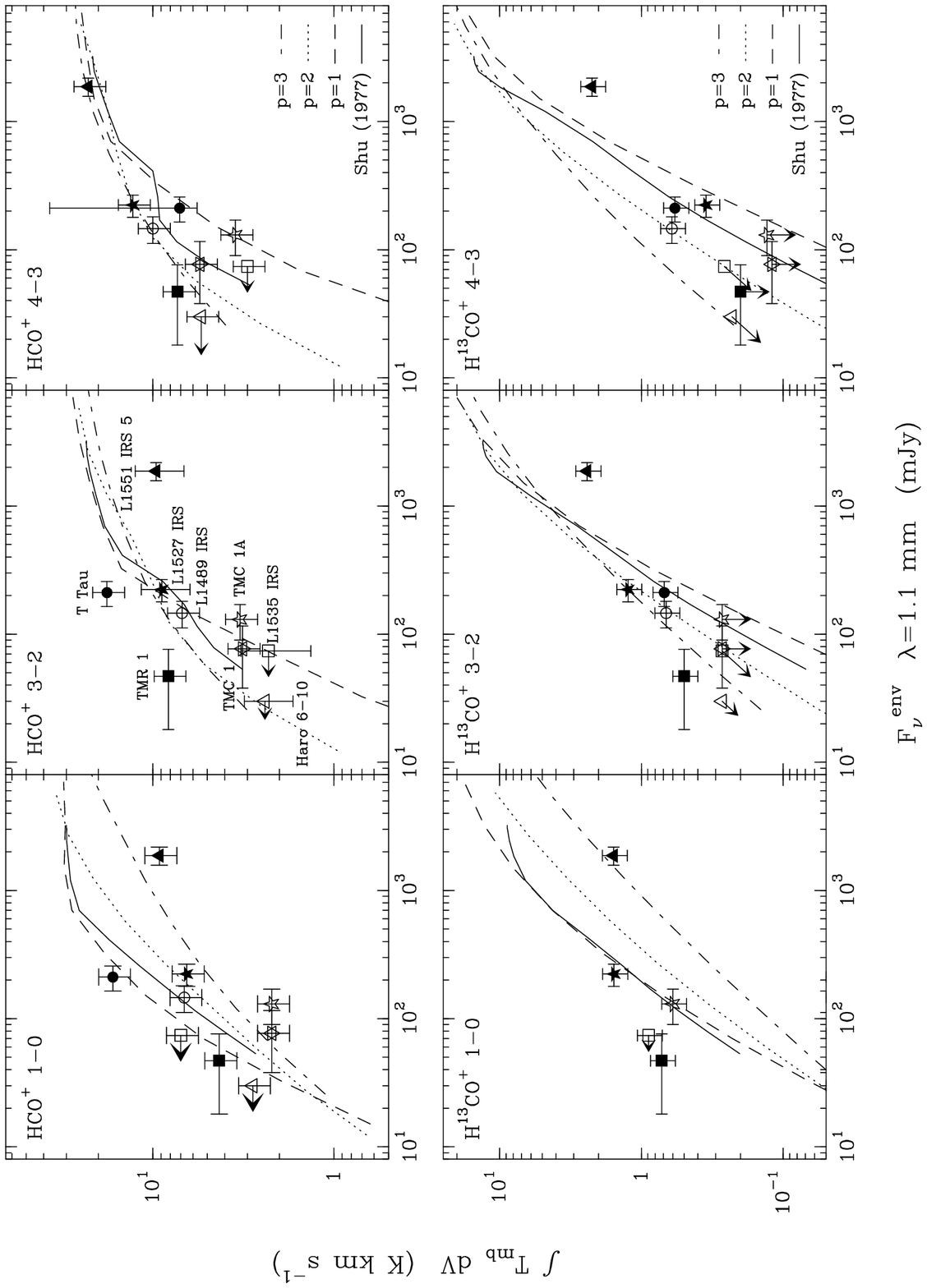,rheight=10truecm,height=25truecm}}
\vfill\eject

\null
\rightline{\tt\underbar{Fig.~6}}
\centerline{\psfig{figure=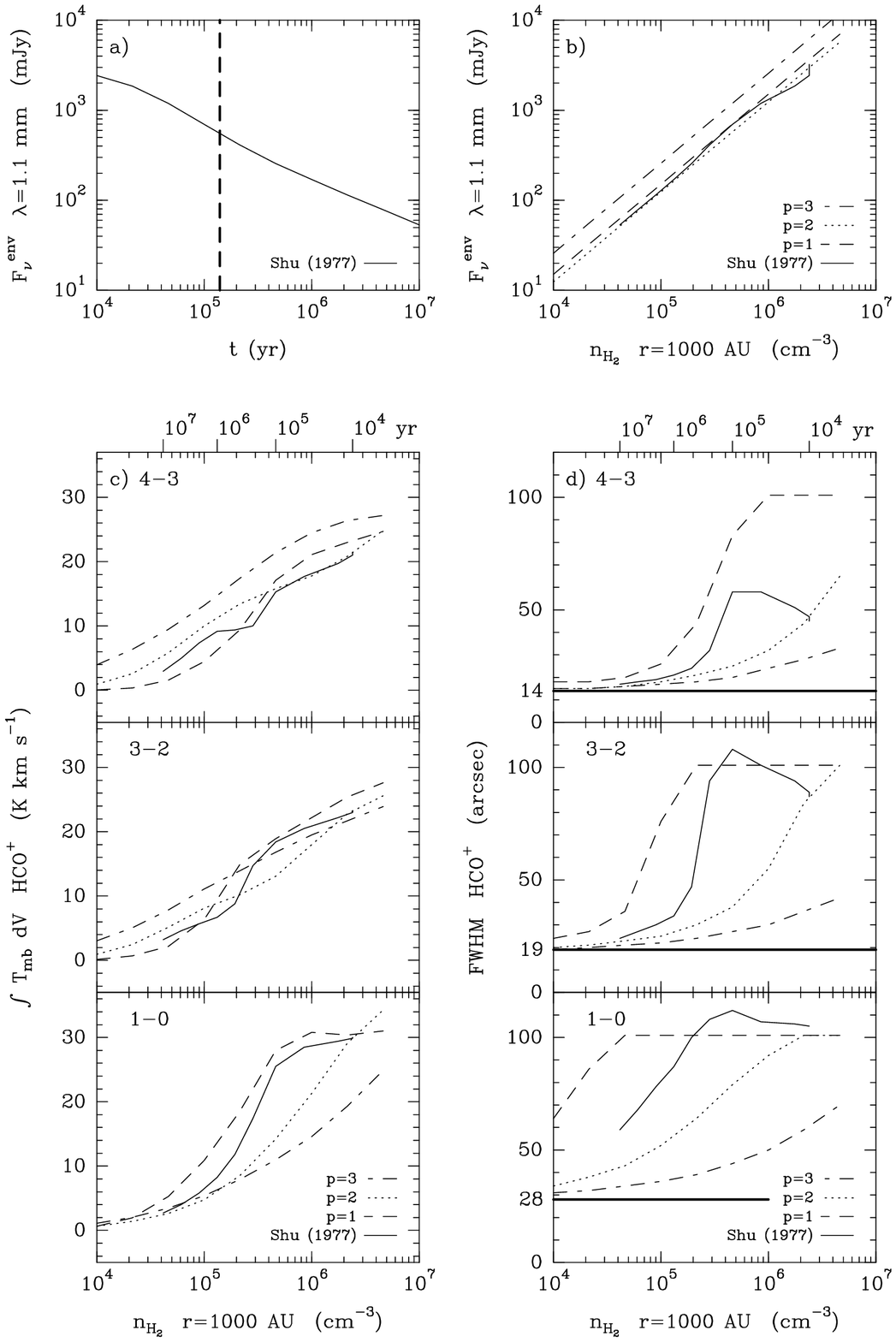,rheight=10truecm,height=25truecm}}
\vfill\eject

\null
\rightline{\tt\underbar{Fig.~7}}
\centerline{\psfig{figure=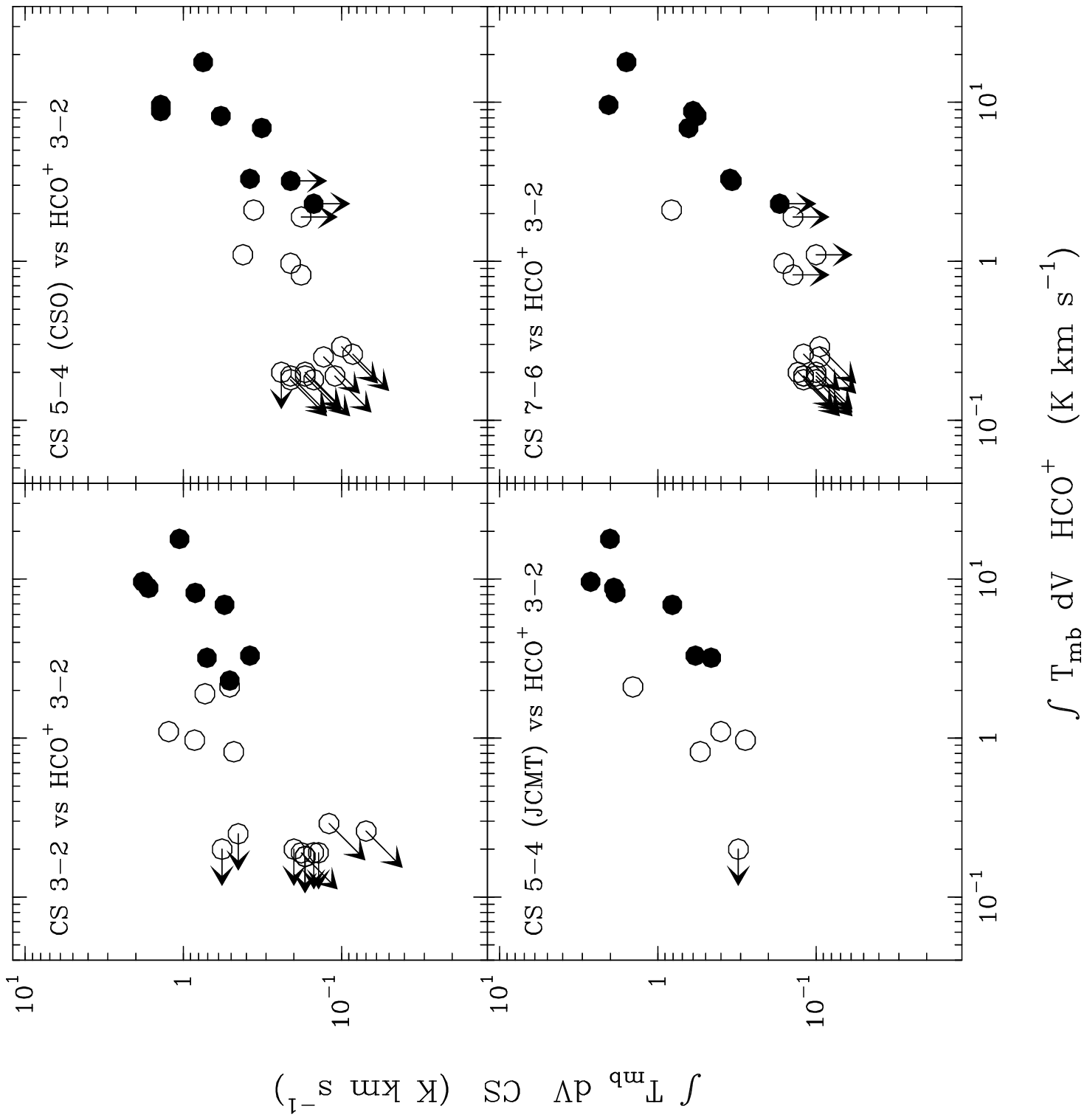,rheight=10truecm,height=25truecm}}
\vfill\eject

\null
\rightline{\tt\underbar{Fig.~8}}
\centerline{\psfig{figure=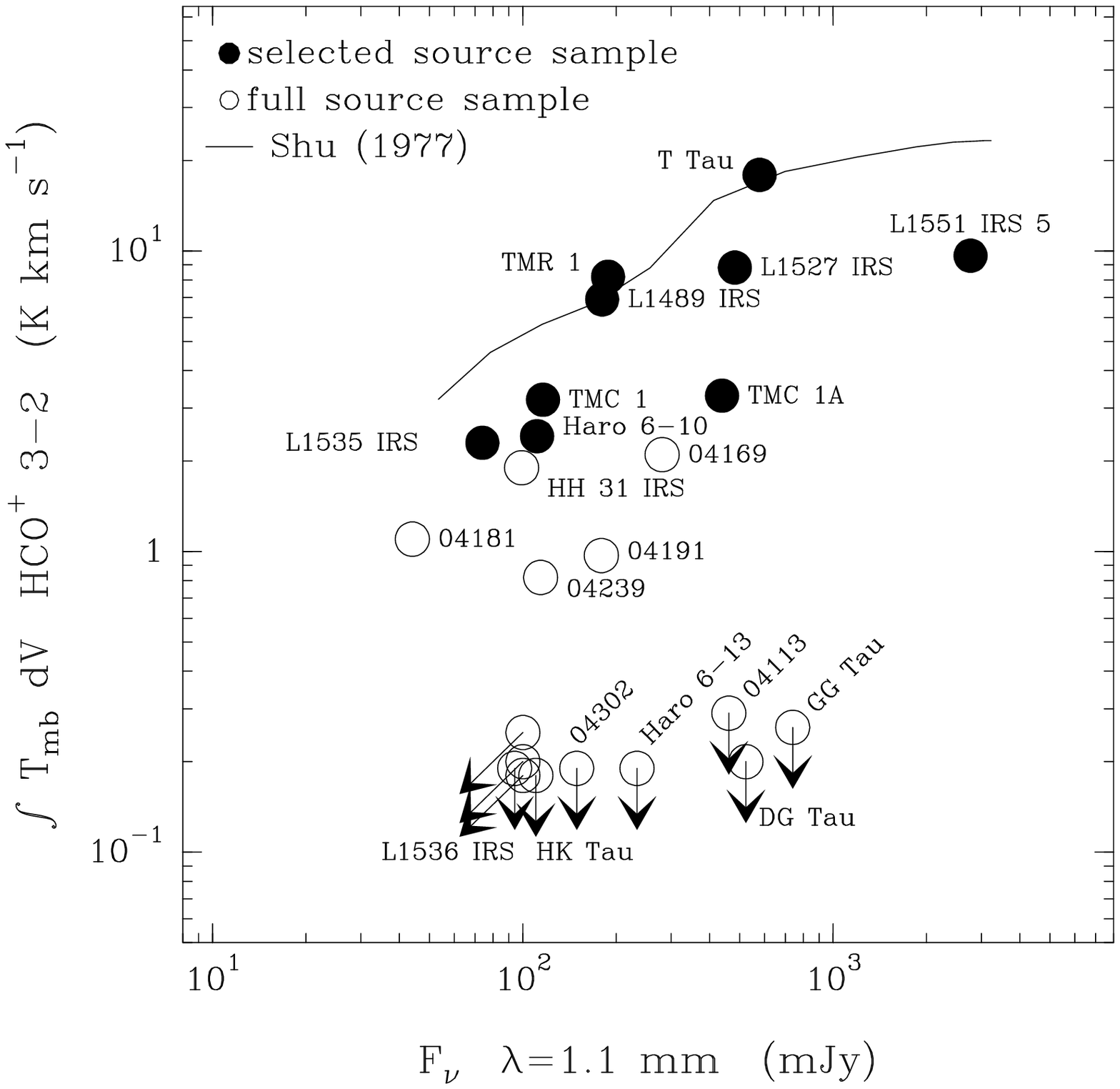,rheight=10truecm,height=25truecm}}
\vfill\eject

\null
\rightline{\tt\underbar{Fig.~9}}
\centerline{\psfig{figure=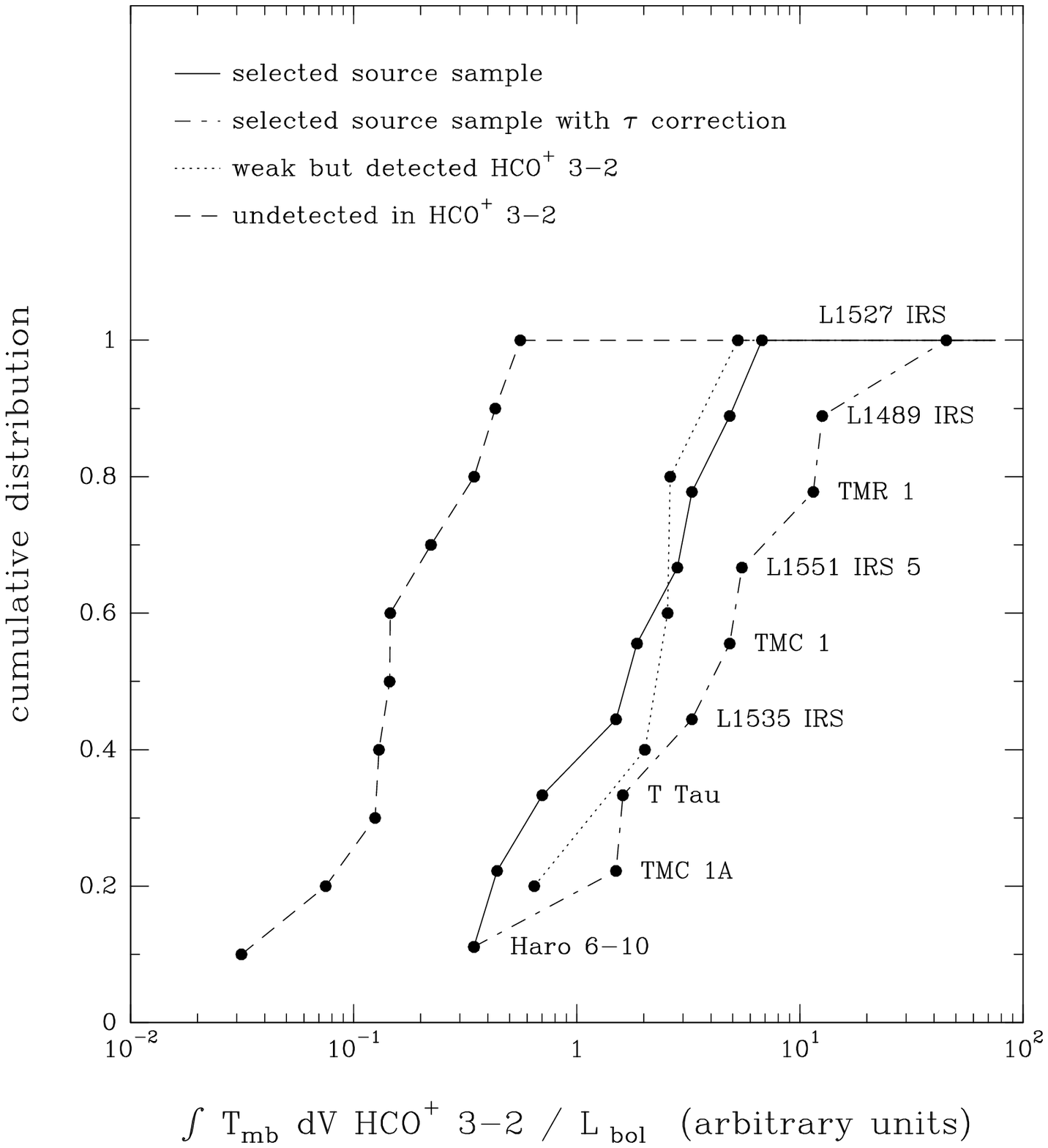,rheight=10truecm,height=25truecm}}
\vfill\eject

\null
\rightline{\tt\underbar{Fig.~10}}
\centerline{\psfig{figure=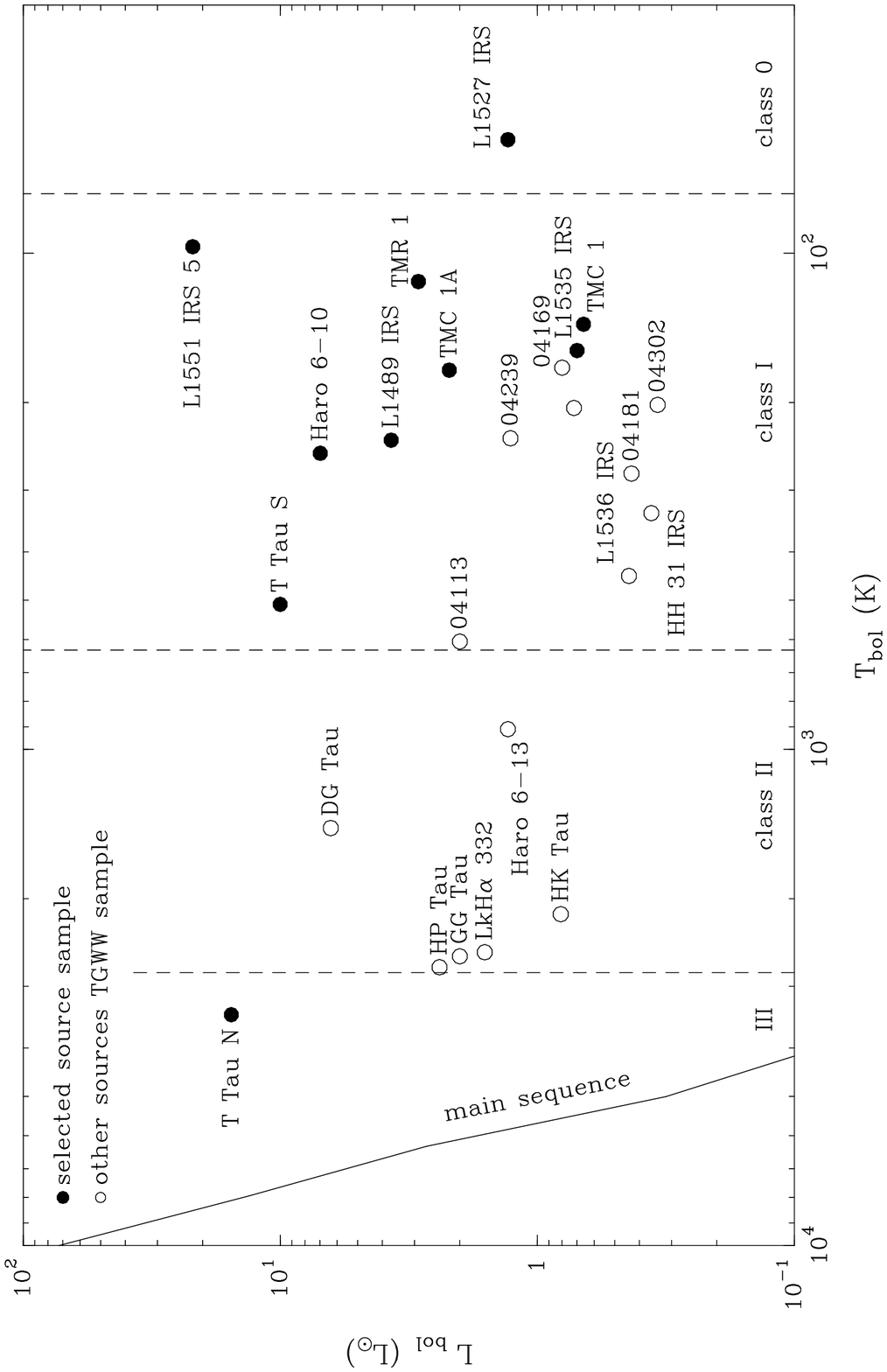,rheight=10truecm,height=25truecm}}
\vfill\eject

\end{document}